\documentclass[pra,aps,english,twocolumn,notitlepage,floatfix]{revtex4-2}
\usepackage{epsfig}
\usepackage{amsmath}
\DeclareMathOperator{\sech}{sech}
\usepackage{mathtools}
\usepackage{xcolor}

\DeclarePairedDelimiter\ket{\lvert}{\rangle}
\DeclarePairedDelimiterX\braket[2]{\langle}{\rangle}{#1 \delimsize\vert #2}
\usepackage{sidecap}
\usepackage[figuresleft]{rotating}
\usepackage{caption}
\usepackage{float}
\usepackage{lipsum}
\usepackage{subfigure}
\usepackage{graphicx}
\usepackage{hyperref}

\begin{document}
 \title{Josephson quantum spin thermodynamics} 
   \author{Subhajit Pal} \author{Colin Benjamin} \email{colin.nano@gmail.com}\affiliation{School of Physical Sciences, National Institute of Science Education \& Research,  An OCC of Homi Bhabha National Institute, Jatni-752050, India }
\begin{abstract} 
A 1D Josephson junction loop, doped with a spin-flipper and attached to two thermal reservoirs is shown to operate as a heat engine, or a refrigerator, or a Joule pump or even as a cold pump. When operating as a quantum heat engine, the efficiency of this device exceeds that of some recent Josephson heat engine proposals. Further, as a quantum refrigerator, the coefficient of performance of this device is much higher than previously proposed Josephson junction based refrigerators. In addition, this device can be tuned from engine mode to refrigerator mode or to any other mode, i.e., Joule pump or cold pump by either tuning the temperature of reservoirs, or via the flux enclosed in the Josephson junction loop. {In presence of spin flip scattering we can tune our device from engine mode to other operating modes by only changing the enclosed flux in Josephson junction loop without changing the temperatures of the reservoirs. This is potentially an advantage with respect to other proposals.} This makes the proposed device much more versatile as regards possible applications.
\end{abstract}
\maketitle
\section{Introduction}
Recently, superconducting hybrid systems have drawn attention due to their possible device applications as sensitive detectors\cite{hei,gua,pso}, low-temperature sensitive thermometers\cite{zgi,wang,luu}, heat valves\cite{yan,soth,dutta}, solid-state quantum machines\cite{mar,vis,kar}, solid-state micro-refrigerators\cite{mml,ngu,bos} and thermoelectric generators\cite{gma,huss,bra}. Quantum thermodynamics implies the study of thermodynamic processes from the principle of quantum mechanics\cite{rko}, while refrigeration means transfer of heat from low to high temperature region\cite{kos} aided by work done on system. In this context  thermodynamic\cite{mjm,est,for,afo} properties of a Josephson Stirling engine have been discussed in Ref.~\onlinecite{bsc}, wherein a quantum spin Hall insulator based Josephson junction is shown to act as a quantum heat engine. Josephson Stirling engines aren't the only game in town, diffusive SNS junctions have been shown to operate as a Josephson-Otto or as Josephson-Stirling engines in Ref.~\onlinecite{carr}.  

In this paper, by doping a spin flipper in a 1D Josephson junction(JJ) loop which is in turn attached to two thermal reservoirs at in-equivalent temperatures via thermal valves, we show that this device can be employed both as a quantum heat engine as well as a refrigerator and can also work as a Joule pump or cold pump even. 

The main advantage our proposed device possesses over other proposals is the tunability by magnetic flux which threads the JJ loop. {In presence of spin flip scattering we can tune our device from engine mode to other operating modes by only changing the enclosed flux in Josephson junction loop without changing the temperatures of the reservoirs. This is potentially an advantage with respect to other proposals}.

{The rest of the paper is arranged as follows. In section II the model is introduced via Hamiltonian, wave functions and boundary conditions so as to calculate different thermodynamic quantities. In section III the thermodynamic processes involved in the Josephson-Stirling cycle are discussed, then work done and heat exchanged during each of these processes is calculated. In section IV the results are {shown} and different operating modes for Josephson-Stirling cycle are discussed. {Next, in section V we give a detailed analysis of our results.} Our paper ends with an experimental realization of the proposed device and the take home messages are aggregated in section VI.}
\section{Theory}
The model device, depicted in Fig.~1, is formed from a 1D superconducting loop\cite{butt} interrupted by a spin flipper. An external magnetic flux $\Phi$ controls the superconducting phase difference across the spin-flipper. The JJ loop is attached, via two thermal valves $v_{L}$ and $v_{R}$, to reservoirs at either end which in turn are at temperatures $T_{L}$ and $T_{R}$. The two reservoirs can exchange heat $Q_{L}$ and $Q_{R}$ with the JJ loop. The scattering problem is solved using BTK approach\cite{BTK} for superconductor-spin flipper-superconductor junction as shown in dashed line box of Fig.~1. The two reservoirs control temperature of JJ loop via thermal valves $v_{L}$ and $v_{R}$. When valve $v_{R}$ is opened and $v_{L}$ is closed, JJ loop is in thermal contact with right reservoir at temperature $T_{R}$. Similarly, when valve $v_{L}$ is opened and $v_{R}$ is closed, JJ loop is in thermal contact with left reservoir at temperature $T_{L}$. On the other hand, phase difference across JJ loop is controlled via magnetic flux $\Phi$ enclosed by the loop. Thus, by controlling both temperature and phase difference, the JJ device can be driven from one state to another. We discuss this in more detail for a Stirling cycle in section III.
\begin{figure}[ht]
\centering{\includegraphics[width=0.4\textwidth]{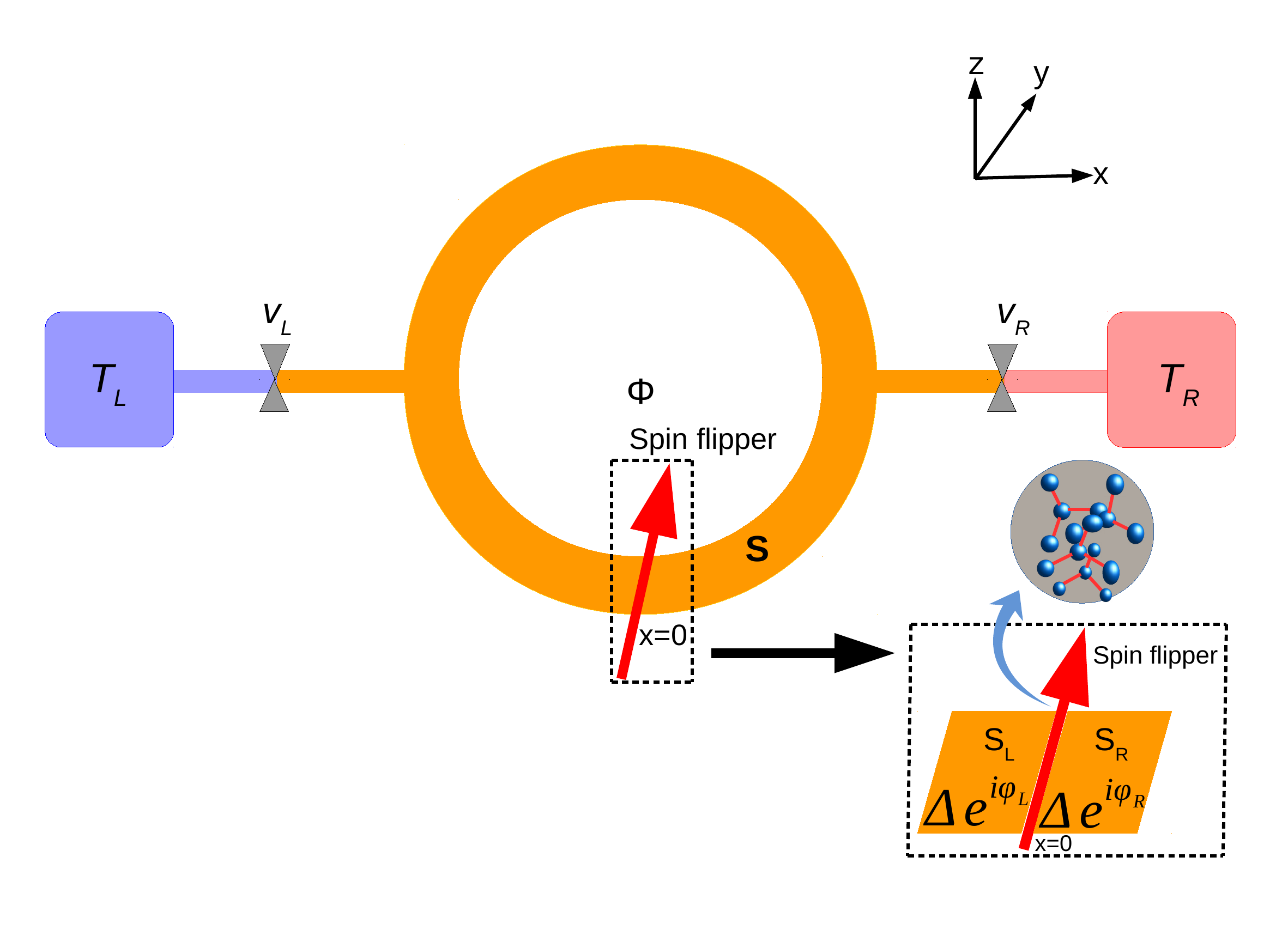}}
\caption{\small \sl 1D Josephson junction (JJ) loop (circumference $L_{S}$, in orange) doped with a spin flipper and attached to two thermal reservoirs at temperatures $T_{L}$ and $T_{R}$ via two thermal valves $v_{L}$ and $v_{R}$. A magnetic flux $\Phi$ controls phase difference $\varphi$ across spin flipper in JJ loop. {High spin molecules like Fe$_{19}$-complex can act as a spin flipper.}}
\end{figure}
\subsection{Hamiltonian} 
In our proposed set-up, a spin-flipper is embedded in the JJ loop of circumference $L_{S}$, see Fig.~1. We used BTK approach\cite{BTK} to solve the scattering problem. {In our work spin-flipper is a delta potential magnetic impurity\cite{AJP} fixed between two superconductors. For proper understanding of our system we compare our delta potential magnetic impurity with a rectangular potential barrier magnetic impurity in Fig.~2. In Fig.~2(a), a single magnetic impurity is lying along the solid black color line at $x=0$. The magnetic impurity is designed as a delta potential along the $x$-direction but is uniform along the $y$-direction. We assume magnetic impurity to have a finite width with a translational invariance along the $y$-direction. Similarly, in Fig.~2(b) we show that a magnetic impurity can have a finite width between $x=0$ and $x=L$ with a translational invariance along the $y$ direction. If we reduce the width $L$ of the impurity, it becomes a delta function like profile influencing the transmission along the $x$ direction but not along the $y$ direction, as shown in Fig.~2(a). Similar concepts have been used to model magnetic impurity in similar junctions, for e.g., graphene-magnetic impurity-graphene junction, see Refs.~\onlinecite{ASC,Maru}.} 
\begin{figure}[ht]
\centering{\includegraphics[width=0.5\textwidth]{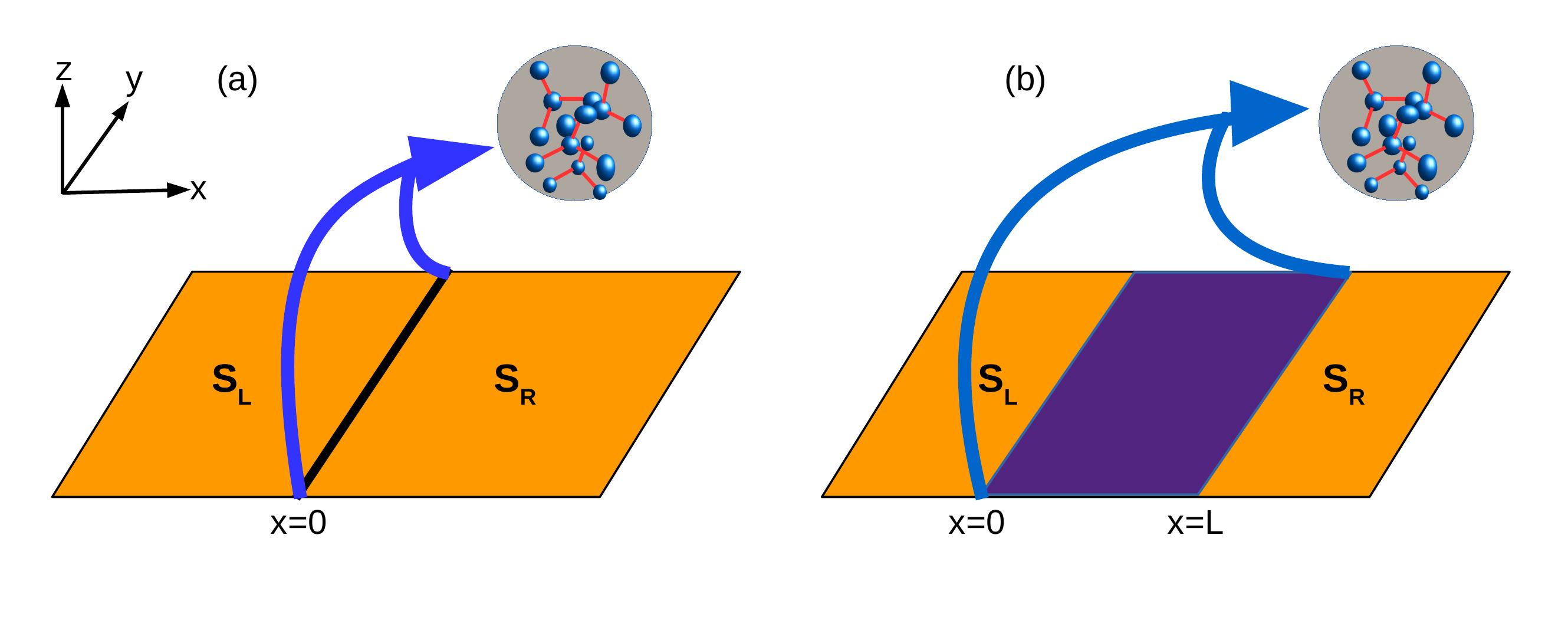}}
\caption{\small \sl Two superconductors separated by (a) a delta potential magnetic impurity, (b) a rectangular barrier magnetic
impurity.}
\end{figure}
Hamiltonian for spin-flipper, from Refs.~\onlinecite{AJP,Liu,Maru,FC,ysr} is,
\begin{equation}
H_{\mbox{spin-flipper}}=-J_{0}\vec{s}.\vec{S},
\label{spinflipper}
\end{equation}
where $J_{0}$ is the strength of exchange interaction, $\vec{s}$ is the spin of electronlike quasiparticle/holelike quasiparticle and $\vec{S}$ is the spin of spin flipper.  
Bogoliubov-de Gennes (BdG) Hamiltonian for the JJ loop is given as\cite{LINDER}
\begin{equation}
H_{BdG}(x)=
\begin{pmatrix}
H\hat{I} & i\Delta(x) \hat{\sigma}_{y}  \\
-i\Delta^{*}(x) \hat{\sigma}_{y}   &  -H\hat{I}
\end{pmatrix},
\label{ham}
\end{equation}
where $H={\nu^2}/2m^{\star}-J_{0}\delta(x)\vec s.\vec S-E_{F}$, with ${\nu^2}/2m^{\star}$ represents kinetic energy of electron-like or hole-like quasiparticle of mass $m^{\star}$ {and momentum $\nu$}, and $E_{F}$ is Fermi energy. Superconducting gap $\Delta(x)$ is of the form $\Delta(x)=\Delta[e^{i\varphi_{L}}\Theta(-x)+e^{i\varphi_{R}}\Theta(x)]$. {For simplicity,} in our work we consider superconducting gap to be independent of temperature, i.e., $\Delta(T)\approx \Delta(T=0)$ with $\frac{\partial\Delta}{\partial T}\approx0$, which is valid if $T\ll T_{c}$, $T_{c}$ being superconducting critical temperature. $\varphi_{L}$ and $\varphi_{R}$ are superconducting phases for left and right superconductors respectively as shown in dashed line box of Fig.~1. In the manuscript, we use dimensionless parameter $J=\frac{m^{\star}J_{0}}{k_{F}}$ as a measure of strength of exchange interaction\cite{AJP} between quasi-particles and spin-flipper.
\subsection{Wavefunctions and boundary conditions}
Diagonalizing BdG Hamiltonian (Eq.~\eqref{ham}) one gets the wavefunctions in superconducting regions of our system for electron/hole like quasiparticle incidence.
For electronlike quasiparticle with spin up incident from left superconductor, wave function for left superconductor is\cite{LINDER}-
\begin{eqnarray}
&&\psi_{S_{L}}(x)=\begin{pmatrix}u\\
                              0\\
                              0\\ 
                              v
                              \end{pmatrix}e^{iq_{+}x}\phi_{m'}^{S}+r_{ee}^{\uparrow\uparrow}\begin{pmatrix}
                              u\\
                              0\\
                              0\\
                              v
                             \end{pmatrix}e^{-iq_{+}x}\phi_{m'}^{S}\nonumber\\&&+r_{ee}^{\uparrow\downarrow}\begin{pmatrix}
                             0\\
                             u\\
                            -v\\
                             0
                             \end{pmatrix}e^{-iq_{+}x}\phi_{m'+1}^{S}+r_{eh}^{\uparrow\uparrow}\begin{pmatrix}
                             0\\
                            -v\\
                             u\\
                             0
                             \end{pmatrix}e^{iq_{-}x}\phi_{m'+1}^{S}\nonumber\\&&+r_{eh}^{\uparrow\downarrow}\begin{pmatrix}
                             v\\
                             0\\
                             0\\
                             u
                             \end{pmatrix}e^{iq_{-}x}\phi_{m'}^{S}, \mbox{ for $x<0$ }.
                             \label{lsi}
                             \end{eqnarray}
$r_{ee}^{\uparrow\uparrow},r_{ee}^{\uparrow\downarrow},r_{eh}^{\uparrow\uparrow},r_{eh}^{\uparrow\downarrow}$ are normal reflection amplitude without any flip, normal reflection amplitude with flip, Andreev reflection amplitude with flip and Andreev reflection amplitude without any flip respectively.\\
The corresponding wave function in right superconductor,
\begin{eqnarray}
\psi_{S_{R}}(x)=&&t_{ee}^{\uparrow\uparrow}\begin{pmatrix}
                              ue^{i\varphi}\\
                              0\\
                              0\\
                              v
                             \end{pmatrix}e^{iq_{+}x}\phi_{m'}^{S}+t_{ee}^{\uparrow\downarrow}\begin{pmatrix}
                             0\\
                             ue^{i\varphi}\\
                             -v\\
                             0
                             \end{pmatrix}e^{iq_{+}x}\phi_{m'+1}^{S}\nonumber\\&&+t_{eh}^{\uparrow\uparrow}\begin{pmatrix}
                             0\\
                             -ve^{i\varphi}\\
                             u\\
                             0
                             \end{pmatrix}e^{-iq_{-}x}\phi_{m'+1}^{S}+t_{eh}^{\uparrow\downarrow}\begin{pmatrix}
                             ve^{i\varphi}\\
                             0\\
                             0\\
                             u
                             \end{pmatrix}e^{-iq_{-}x}\nonumber\\&&\phi_{m'}^{S},\mbox{ for $x>0$ },
                             \label{rsi}
                             \end{eqnarray}
$t_{ee}^{\uparrow\uparrow},t_{ee}^{\uparrow\downarrow},t_{eh}^{\uparrow\uparrow},t_{eh}^{\uparrow\downarrow}$ being transmission amplitudes, corresponding to reflection processes described above and $\varphi=\varphi_{R}-\varphi_{L}$ is phase difference between right and left superconductors. $\phi_{m'}^{S}$ represents eigenspinor of spin flipper, with the $S^{z}$ operator of spin-flipper acting as, $S^{z}\phi_{m'}^{S} = m'\phi_{m'}^{S}$, $m'$ being spin magnetic moment of spin-flipper. BCS coherence factors are $u=\sqrt{\frac{1}{2}\Bigg(1+\frac{\sqrt{E^2-\Delta^{2}}}{E}\Bigg)}$, $v=\sqrt{\frac{1}{2}\Bigg(1-\frac{\sqrt{E^2-\Delta^{2}}}{E}\Bigg)}$. Wavevectors for electron-like quasiparticles ($q_{+}$) and hole-like quasiparticles ($q_{-}$) are 
$q_{\pm}=\sqrt{\frac{2m^{\star}}{\hbar^2}(E_{F}\pm \sqrt{E^2-\Delta^2})}$. Andreev approximation\cite{Kri} gives $q_{+}=q_{-}=k_{F}$, with $k_{F}$ being Fermi wavevector, and $E_{F}\gg\Delta$. Imposing boundary conditions on (\ref{lsi}, \ref{rsi}) at $x=0$, gives\\
\begin{equation}
\psi_{S_{L}}(x)=\psi_{S_{R}}(x),\,\, \frac{d\psi_{S_{R}}}{dx}-\frac{d\psi_{S_{L}}}{dx}=-\frac{2m^{\star}J_{0}\vec s.\vec S}{\hbar^2} \psi_{S_{L}},
\label{bc}
\end{equation}
where $\vec s.\vec S=s^{z}S^{z}+\frac{1}{2}(s^{-}S^{+}+s^{+}S^{-})$ represents exchange operator in Eq.~\eqref{spinflipper}, with
$s^{\pm} = s_{x}\pm is_{y}$ and $S^{\pm} = S_{x}\pm iS_{y}$ are spin raising and spin lowering operators for electronlike quasiparticle/holelike quasiparticle and spin-flipper respectively. {In our theoretical treatment, we solve the scattering problem using BTK approach for superconductor-spin flipper-superconductor junction as shown in dashed line box of Fig.~1. As depicted in Fig.~1, spin flipper is placed at $x=0$ and there is a phase difference $\varphi$ across the spin flipper. This phase difference is generated by magnetic flux $\Phi$ in Josephson junction loop, which can control it. We solve the scattering problem at $x=0$, thus our results do not depend on $x$ or loop circumference $L_{S}$.} In Appendix A we give a detailed calculation as to how spin flipper interacts with electron-like and hole-like quasiparticles.
\subsection{Andreev bound states}
To calculate Andreev bound states (see Ref.~\cite{annu} for details of the method) we neglect contribution from incoming quasiparticle and insert wavefunctions (\ref{lsi}, \ref{rsi}) into boundary conditions (\ref{bc}). We get a homogeneous system of $8$ linear equations for the scattering amplitudes,
\begin{equation}
Py=0 
\end{equation}
where $y$ is a $8\times1$ column matrix and given as $y=[r_{ee}^{\uparrow\uparrow},r_{ee}^{\uparrow\downarrow},r_{eh}^{\uparrow\uparrow},r_{eh}^{\uparrow\downarrow},t_{ee}^{\uparrow\uparrow},t_{ee}^{\uparrow\downarrow},t_{eh}^{\uparrow\uparrow},t_{eh}^{\uparrow\downarrow}]$, $P$ being a $8\times8$ matrix. For nontrivial solution of this system, the determinant of $P=0$ and we get Andreev bound states as a function of phase difference $\varphi$ between two superconductors, i.e., Andreev bound state energy spectrum $E_{j}$, $j=\{1,...,4\}$\cite{Been}. We find that $E_{j}(\varphi)=E_{\sigma}^{\pm}(\varphi)=\pm E_{\sigma}(\varphi), (\sigma=\uparrow,\downarrow)$ and 
\begin{equation}
E_{\sigma}^{\pm}(\varphi)=\pm\Delta\sqrt{1+\frac{A(\varphi)}{C}+\rho_{\sigma}\frac{\sqrt{B(\varphi)}}{C}},\label{Eq:eb} 
\end{equation}
\begin{eqnarray}
\mbox{ where, } A(\varphi)=&&J^2(2+p^4J^2+2p^2(-2+J^2m'(1+m'))\nonumber\\
                           &&+m'(1+m')(4+J^2m'(1+m')))\nonumber\\
                           &&+2(8+J^2(1-2p^2+2m'(1+m')))\cos(\varphi),\nonumber
\end{eqnarray}
\begin{eqnarray}
B(\varphi)=&&2J^2(64p^4J^2+3(J+2Jm')^2+4p^2\nonumber\\
                           &&(16+J^2(5+4m'(1+m')))+4J^2(-4p^2+16p^4-\nonumber\\
                           &&(1+2m')^2)\cos(\varphi)+((J+2Jm')^2-4p^2(16+\nonumber\\
                           &&(J+2Jm')^2))\cos(2\varphi)),\nonumber\\
C=(1&&6+J^4(p^2+m'+m'^2)^2+J^2(4+8p^2+8m'(1+m'))),\nonumber
\end{eqnarray}
$\rho_{\uparrow(\downarrow)}=+1(-1)$, {and $p=\sqrt{(S-m')(S+m'+1)}$ is the spin flip probability\cite{AJP} for spin flipper, {where $S$ is the spin flipper's spin and $m'$ is the spin flipper's spin magnetic moment.}\\
In absence of spin flip scattering ($p=0$), Eq.~\eqref{Eq:eb} reduces to, 
\begin{equation}
\label{ebn}
E(\varphi)=\pm\Delta\sqrt{\frac{4\cos^2(\varphi/2)+J^2m'^2}{4+J^2m'^2}}. 
\end{equation}
{In the absence of spin flipper ($J=0$), Eq.~\eqref{ebn} reduces to the well known result\cite{kulik}
\begin{equation}
E(\varphi)=\pm\Delta\cos(\varphi/2). 
\end{equation}}
{ In Fig.~3 we plot Andreev bound states as function of phase difference $\varphi$ for different values of flip probability of spin flipper. We see that in absence of spin flip scattering ($p=0$), there are two bound states, however in presence of spin flip scattering ($p\neq0$) number of bound states increase to four. We also notice that for no flip process, two Andreev bound states touch at $\varphi=\pi$. However, in presence of spin flip scattering Andreev bound states no longer touch at $\varphi=\pi$ but are shifted with respect to it. For $p=3$, two Andreev bound state energies touch at $\varphi=0.91\pi$ and $\varphi=1.09\pi$, while for $p=5$ the two Andreev bound state energies touch at $\varphi=0.84\pi$ and $\varphi=1.16\pi$. In absence of spin flip scattering Andreev bound states are degenerate at $\varphi=\pi$, spin flip scattering lifts this degeneracy.}
\begin{figure}[ht]
\centering{\includegraphics[width=0.5\textwidth]{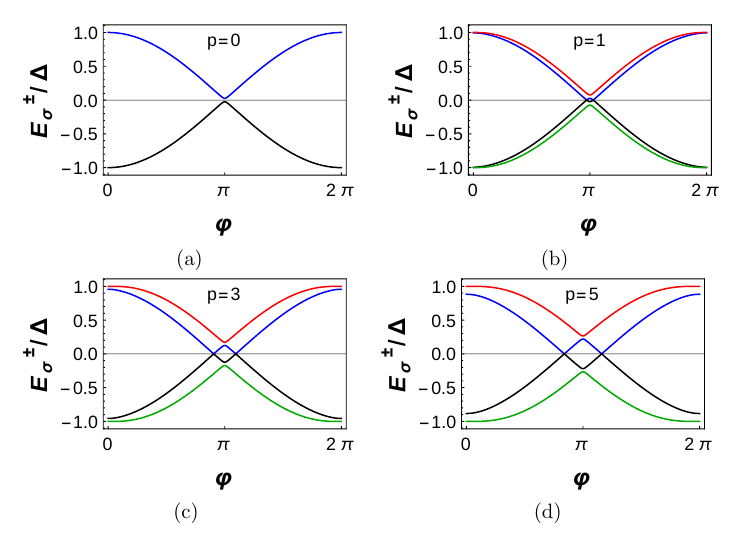}}
\caption{\small \sl {Andreev bound states as function of phase difference $\varphi$ for different values of flip probability $p$ of spin flipper. Parameters: $J=0.1$, (a) $S=m'=1/2$, $p=0$; (b) $S=1/2$, $m'=-1/2$, $p=1$; (c) $S=5/2$, $m'=-1/2$, $p=3$; (d) $S=9/2$, $m'=-1/2$, $p=5$.}}
\end{figure}
\subsection{Superconductor-Spin flipper-Superconductor junction as a thermodynamic system}
Before analyzing the thermodynamic behavior of setup (Fig.~1), we introduce the different thermodynamic quantities. From Andreev bound state energies we can determine {phase-dependent part of} Free energy of JJ\cite{golu} (inset of Fig.~1) as
\begin{eqnarray}
F(\varphi,T)&&=-\frac{1}{\beta}\ln\Big[\prod_{j}(1+e^{-\frac{E_{j}(\varphi)}{k_{B}T}})\Big],\nonumber\\
            &&= -\frac{2}{\beta}\sum_{\sigma}\ln\Big[2\cosh\Big(\frac{E_{\sigma}(\varphi)}{2k_{B}T}\Big)\Big],
\label{ff}
\end{eqnarray}
where $k_{B}$ is Boltzmann constant. {In Eq.~\eqref{ff}, we neglect the contribution from the quasiparticle states in the continuum with energies above the superconducting gap, whose density of states is $\rho_{c}$. In case of short Josephson junction (length of the weak link much smaller than the superconducting coherence length) $\rho_{c}$ is same as in a bulk superconductor and therefore is phase independent. For Superconductor-Spin flipper-Superconductor junction across the spin flipper, the junction length is infinitesimally small and thus $\rho_{c}$ is phase independent. The phase-independent part of Free energy does not contribute to Josephson current, work done and the heat exchanged.} From Free energy we can calculate total Josephson current\cite{deG} as
\begin{equation}
I(\varphi,T)=\frac{2e}{\hbar}\frac{\partial F(\varphi,T)}{\partial \varphi},
\label{ff1}
\end{equation}
where $e$ is charge of electron. Entropy of our device can then be calculated from Free energy as
\begin{equation}
\Omega(\varphi,T)=-\frac{\partial F(\varphi,T) }{\partial T}.
\end{equation}
From entropy $\Omega$ one determines heat capacity of JJ as
\begin{equation}
C(\varphi,T)=T\frac{\partial \Omega(\varphi,T)}{\partial T}. 
\end{equation}
\section{Josephson-Stirling Cycle:}
Josephson-Stirling cycle\cite{carr,bsc} represented in Fig.~4, involves two isothermal and two isophasic processes. States $\textbf1$,$\textbf{2}$ involve right reservoir, while states $\textbf3$,$\textbf{4}$ involve left reservoir. Below we summarize these different processes-
\begin{itemize}
 \item \textbf{Isothermal process ($\textbf{1}\rightarrow\textbf{2}$)}: Thermal valve $v_{R}$ is open while $v_{L}$ is closed, thus system is in thermal contact with right reservoir at temperature $T_{R}$. The device or system  goes from state $\textbf{1}\equiv(\varphi=0,T_{R})$ to state $\textbf{2}\equiv(\varphi=\varphi_{f},T_{R})$.
 \item \textbf{Isophasic process ($\textbf{2}\rightarrow\textbf{3}$)}: Thermal valve $v_{L}$ is open while $v_{R}$ is closed and system is driven from state $\textbf{2}\equiv(\varphi=\varphi_{f},T_{R})$ to state $\textbf{3}\equiv(\varphi=\varphi_{f},T_{L})$.
 \item \textbf{Isothermal process ($\textbf{3}\rightarrow\textbf{4}$)}: In this stage valve $v_{R}$ is closed and $v_{L}$ is open. System is transferred from state $\textbf{3}\equiv(\varphi=\varphi_{f},T_{L})$ to state $\textbf{4}\equiv(\varphi=0,T_{L})$.
 \item \textbf{Isophasic process ($\textbf{4}\rightarrow\textbf{1}$)}: Final stage of cycle involves closing valve $v_{L}$ and opening valve $v_{R}$, with system being driven from state $\textbf{4}\equiv(\varphi=0,T_{L})$ to state $\textbf{1}\equiv(\varphi=0,T_{R})$.
\end{itemize}
\begin{figure}[ht]
\centering{\includegraphics[width=.75\linewidth]{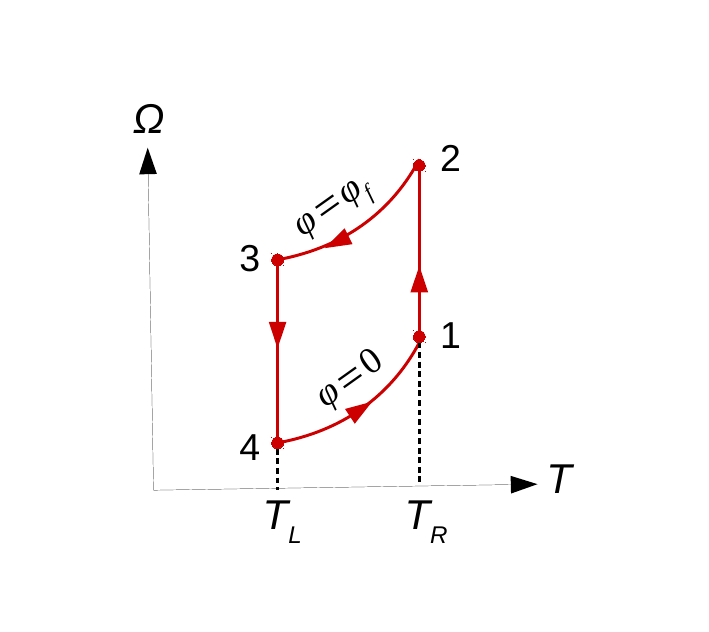}}
\caption{\small \sl Josephson-Stirling cycle in $T\Omega$ ($T$ is temperature and $\Omega$ is entropy) plane. Enclosed area in $T-\Omega$ plane corresponds to  total heat exchanged $Q$, which is equal to total work done $W$ during the cycle.}
\end{figure}
{When the device is driven from state $i$ to state $f$ ($i\rightarrow f$) during a quasi-static process, work done and heat released during the process are given as $W_{if}=-\frac{\hbar}{2e}\int_{\varphi_{i}}^{\varphi_{f}} I(\varphi,T)d\varphi$ and $Q_{if}=\int_{\Omega(\varphi_{i},T)}^{\Omega(\varphi_{f},T)} T d\Omega$ respectively, where $I(\varphi,T)$ is Josephson current and $\Omega$ is entropy of setup depicted in Fig.~1. In expressions for work done and heat released below, we have taken sign convention such that $W_{if}$ is positive when work is released to the universe while $Q_{if}$ is positive when heat is absorbed from the universe.
Work done and heat released for an isothermal process, where phase difference $\varphi$ changes from $\varphi_{i}\rightarrow\varphi_{f}$ at constant temperature $T$ is given as $W_{if}=-[F(\varphi_{f},T)-F(\varphi_{i},T)]$ and $Q_{if}=T[\Omega(\varphi_{f},T)-\Omega(\varphi_{i},T)]$ respectively. For an isophasic process, temperature changes from $T_{i}\rightarrow T_{f}$ at constant phase difference $\varphi$, $W_{if}=0$ and $Q_{if}=\int_{T_{i}}^{T_{f}}C(\varphi,T) dT$.}
\subsection{Work done and heat exchanged in Josephson-Stirling Cycle}
We can now explicitly calculate total work done and heat exchanged during each stage of the Josephson-Stirling cycle, shown in Fig.~4. We also distinguish between four distinct modes of operation of the Josephson-Stirling cycle. {In heat engine mode: $W>0$, $Q_{R}>0$ and $Q_{L}<0$, where $W$ is the work done, $Q_{R}$ and $Q_{L}$ are the heat exchanged with right and left reservoirs respectively. This implies when the Josephson-Stirling cycle operates as an engine, work is done by the system on the universe, the cycle absorbs heat $Q_{R}$ from the hot reservoir with temperature $T_{R}$ and releases heat $|Q_{L}|$ ($<Q_{R}$) to the cold reservoir with temperature $T_{L}$. In refrigerator mode: $W<0$, $Q_{R}>0$ and $Q_{L}<0$. Thus, when the cycle acts as a refrigerator, work is done on the system by the universe, the cycle absorbs heat $Q_{R}$ from the cold reservoir with temperature $T_{R}$ and releases heat $|Q_{L}|$ ($>Q_{R}$) to the hot reservoir with temperature $T_{L}$. Further in Joule pump mode: $W<0$, $Q_{R}<0$ and $Q_{L}<0$. Thus, when the cycle acts as a Joule pump, it completely converts work into heat released to the reservoirs. Finally, in cold pump mode: $W<0$, $Q_{R}<0$ and $Q_{L}>0$. This implies when the cycle operates as a cold pump, it absorbs heat $Q_{L}$ from the hot reservoir with temperature $T_{L}$ and releases heat $|Q_{R}|$ to the cold reservoir with temperature $T_{R}$.}
{\subsubsection{Work done}}
The total work done per cycle is $W=W_{12}+W_{34}$. Since $2\rightarrow3$ and $4\rightarrow1$ are isophasic processes, therefore $W_{23}=W_{41}=0$. Thus, work done $W$, can be calculated as,
\begin{widetext}
\footnotesize{
\begin{eqnarray}
W=&&\Delta\Bigg\{\Bigg(y_{R}\ln\Bigg[4\cosh\Bigg(\frac{x_{R} E_{\uparrow}(\varphi_{f})}{\Delta}\Bigg)\Bigg]+y_{R}\ln\Bigg[\cosh\Bigg(\frac{x_{R} E_{\downarrow}(\varphi_{f})}{\Delta}\Bigg)\Bigg]-y_{L}\ln\Bigg[4\cosh\Bigg(\frac{x_{L} E_{\uparrow}(\varphi_{f})}{\Delta}\Bigg)\Bigg]-y_{L}\ln\Bigg[\cosh\Bigg(\frac{x_{L} E_{\downarrow}(\varphi_{f})}{\Delta}\Bigg)\Bigg]\Bigg)\nonumber\\
&&-\Bigg(y_{R}\ln\Big[\cosh\Big(x_{R}\Big)\Big]+y_{R}\ln\Big[4\cosh\Big(x_{R} C^{\prime}\Big)\Big]-y_{L}\ln\Big[\cosh\Big(x_{L}\Big)\Big]-y_{L}\ln\Big[4\cosh\Big(x_{L}C^{\prime}\Big)\Big]\Bigg)\Bigg\},\label{work}\nonumber\\\\
\mbox{ where, }&&x_{L}=\frac{\Delta}{2k_{B}T_{L}},\,\, x_{R}=\frac{\Delta}{2k_{B}T_{R}},\,\, y_{L}=\frac{1}{x_{L}},\,\, y_{R}=\frac{1}{x_{R}},\,\, C_{1}=\sqrt{16+J^4 (p^2+m'+m'^2)^2+J^2 (4-8p^2+8m'(1+m'))},\nonumber\\
C_{2}=&&\sqrt{16+J^4(p^2+m'+m'^2)^2+J^2(4+8p^2+8m'(1+m'))},\, \mbox{ and } C^{\prime}=\frac{C_{1}}{C_{2}}.\nonumber 
\end{eqnarray}}
\normalsize In absence of spin flip scattering ($p=0$), work done, \textquoteleft$W$\textquoteright{} in Eq.~\eqref{work}, reduces to,
\begin{equation}
W=\Delta\Bigg\{\Bigg(y_{L}\ln\Big[\cosh\Big(x_{L}\Big)\Big]-y_{R}\ln\Big[\cosh\Big(x_{R}\Big)\Big]\Bigg)-\Bigg(y_{L}\ln\Bigg[\cosh\Bigg(\frac{x_{L}E(\varphi_{f})}{\Delta}\Bigg)\Bigg]-y_{R}\ln\Bigg[\cosh\Bigg(\frac{x_{R}E(\varphi_{f})}{\Delta}\Bigg)\Bigg]\Bigg)\Bigg\},\label{worknoflip}
\end{equation}
where, $E(\varphi)=\Delta\sqrt{\frac{4\cos^2(\varphi/2)+J^2m'^2}{4+J^2m'^2}}$ is Andreev bound state energy for no flip case. 
{\subsubsection{Heat exchanged (with right reservoir)}}
The heat exchanged with right reservoir at temperature $T_{R}$, $Q_{R}=Q_{12}+Q_{41}$ is
\begin{eqnarray}
Q_{R}=&&\Delta\Bigg(\ln\Bigg[\cosh\Bigg(\frac{x_{R} E_{\uparrow}(\varphi_{f})}{\Delta}\Bigg)\Bigg]+\ln\Bigg[4\cosh\Bigg(\frac{x_{R} E_{\downarrow}(\varphi_{f})}{\Delta}\Bigg)\Bigg]\Bigg)y_{R}-E_{\uparrow}(\varphi_{f})\tanh\Bigg(\frac{x_{R}E_{\uparrow}(\varphi_{f})}{\Delta}\Bigg)\nonumber\\  
     &&-E_{\downarrow}(\varphi_{f})\tanh\Bigg(\frac{x_{R}E_{\downarrow}(\varphi_{f})}{\Delta}\Bigg)+\Delta C^{\prime}\tanh\Big(x_{L} C^{\prime}\Big)+\Delta\tanh\Big(x_{L}\Big)
     -\Delta\Bigg(\ln\Big[\cosh\Big(x_{R}\Big)\Big]+\ln\Big[4\cosh\Big(x_{R} C^{\prime}\Big)\Big]\Bigg)y_{R}.\nonumber\\
\label{hrr}
\end{eqnarray}
For no flip process ($p=0$), Eq.~\eqref{hrr} reduces to,
\begin{equation}
\label{QRnoflip}
Q_{R}=\Delta\Bigg(-\ln\Big[\cosh\Big(x_{R}\Big)\Big]+\ln\Bigg[\cosh\Bigg(\frac{x_{R}E(\varphi_{f})}{\Delta}\Bigg)\Bigg]\Bigg)y_{R}+\Delta\tanh\Big(x_{L}\Big)-E(\varphi_{f})\tanh\Bigg(\frac{x_{R}E(\varphi_{f})}{\Delta}\Bigg).
\end{equation}
{\subsubsection{Heat exchanged (with left reservoir)}}
Finally, heat exchanged with left reservoir at temperature $T_{L}$, $Q_{L}=Q_{23}+Q_{34}$ is
\begin{eqnarray}
Q_{L}=&&E_{\uparrow}(\varphi_{f})\tanh\Bigg(\frac{x_{R}E_{\uparrow}(\varphi_{f})}{\Delta}\Bigg)+E_{\downarrow}(\varphi_{f})\tanh\Bigg(\frac{x_{R}E_{\downarrow}(\varphi_{f})}{\Delta}\Bigg)+\Delta\Bigg(\ln\Big[\cosh\Big(x_{L} C^{\prime}\Big)\Big]+\ln\Big[4\cosh\Big(x_{L}\Big)\Big]\Bigg)y_{L}\nonumber\\
     &&-\Delta C^{\prime}\tanh\Big(x_{L} C^{\prime}\Big)-\Delta\tanh\Big(x_{L}\Big)-\Delta\Bigg(\ln\Bigg[4\cosh\Bigg(\frac{x_{L} E_{\uparrow}(\varphi_{f})}{\Delta}\Bigg)\Bigg]+\ln\Bigg[\cosh\Bigg(\frac{x_{L} E_{\downarrow}(\varphi_{f})}{\Delta}\Bigg)\Bigg]\Bigg)y_{L}.\nonumber\\
\label{hlr}
\end{eqnarray}
For no flip process ($p=0$), Eq.~\eqref{hlr}, reduces to,
\begin{equation}
\label{QLnoflip}
Q_{L}=\Delta\Bigg(\ln\Big[\cosh\Big(x_{L}\Big)\Big]-\ln\Bigg[\cosh\Bigg(\frac{x_{L}E(\varphi_{f})}{\Delta}\Bigg)\Bigg]\Bigg)y_{L}-\Delta\tanh\Big(x_{L}\Big)+E(\varphi_{f})\tanh\Bigg(\frac{x_{R}E(\varphi_{f})}{\Delta}\Bigg). 
\end{equation}
\end{widetext}

{Upper limit for $T_{L}$ and $T_{R}$ is much smaller than the superconducting transition temperature $T_{c}$.} For \textquoteleft Pb\textquoteright{} superconductor, $T_{c}$ is $7$K, therefore upper limit for $T_{R}$ and $T_{L}$ is taken as {$3.7$K.} In Figs.~4-9, we consider lead ($Pb$) superconductor, thus upper limit for $T_{R}$, $T_{L}$ is fixed at {$3.7$K}. From conservation of energy $W=Q$, with $Q=Q_{R}+Q_{L}$ being total heat exchanged during the cycle. When Josephson-Stirling cycle acts as a heat engine, work done $W$ and efficiency of cycle is $\eta=W/Q_{R}$. In refrigerator mode, the work done $W$ and coefficient of performance\cite{carr} of Josephson-Stirling cycle is given as $\mbox{COP}=Q_{R}/|W|$. Finally, in Joule pump mode the $COP$ is\cite{biz} $(|Q_{L}|+|Q_{R}|)/|W|$ with {$Q_R<0$ and $Q_L<0$,} as Joule pump converts the work to heat and transfers it to both reservoirs. A cold pump, on the other hand, has $COP=|Q_R|/|W|$\cite{biz}, as in cold pump $Q_R$ is negative implying heat energy is released to right reservoir which is cold reservoir because $T_L>T_R$. A cold pump is used to heat the colder reservoir by transferring heat from hotter reservoir in turn cooling the hotter reservoir. Please also note that the expressions for COP in case of Joule pump  differs slightly from that in Ref.~\cite{biz} as we have taken the sign convention of heat flow or work done as in Ref.~\cite{bsc}.
\section{Results}
\subsection{Josephson-Stirling cycle as quantum heat engine} 
In Fig.~5 we plot work done $W$ (Eq.~\eqref{work}) and efficiency $\eta$ of Josephson quantum heat engine and compare no flip ({flip probability $p=0$, {i.e., $S=m'$ where $S$ is the spin and $m'$ is the spin magnetic moment of the spin-flipper,} see Appendix A}) with spin flip ($p\neq0$, {i.e., $S\neq m'$}) processes. In Figs.~5(a) and 5(c), $W$ and $\eta$ are plotted as function of maximal phase change $\varphi_{f}$ during the cycle. The reason we plot work done {and} efficiency as function of phase is because phase lends itself to external control via magnetic flux enclosed in Josephson junction loop, see Fig.~1. We consider here right reservoir to be hot and left reservoir to be cold, i.e., $T_{R}>T_{L}$. 
\begin{figure}[ht]
\centering{\includegraphics[width=.99\linewidth]{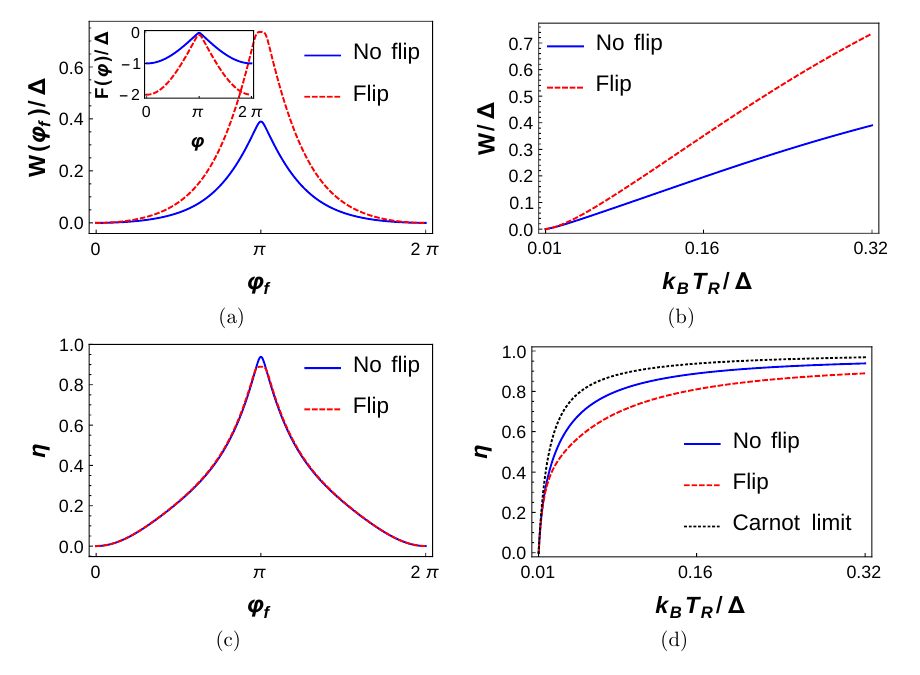}}
\caption{\small \sl Total work done $W$ and efficiency $\eta$ of a Josephson-Stirling engine as a function of $\varphi_{f}$ (for (a,c)) and as a function of the right reservoir temperature $T_{R}$ (for (b,d)). Parameters: $J=0.1$, $\varphi_{i}=0$, $\varphi_{f}=\pi$ (for (b,d)), {$k_{B}T_{R}=0.32\Delta$} (for (a,c)), $k_{B}T_{L}=0.01\Delta$, $k_{B}T=0.01\Delta$, {where $T$ is the system temperature.} The black dotted line represents Carnot limit, $\eta_{C}=1-\frac{T_{L}}{T_{R}}$, Flip: $S=-m'=1/2$, $p=1$; No Flip: $S=m'=1/2$, $p=0$.}
\end{figure}
In the inset of Fig.~5(a), we show Free energy as function of $\varphi$. {We consider system temperature $T$ to be $0.12$K, which is much smaller than the superconducting critical temperature $T_{c}$.}
In Fig.~5(a) we see that $W$ is maximum at $\varphi_{f}=\pi$, irrespective of spin-flip scattering, however magnitude of $W_{max}$ for no flip process is much smaller than spin-flip process. 
{The total work done is maximum at $\varphi_f=\pi$, thus the maximum work done in a cycle is $W_{max}=F(0,T_{R})-F(\pi,T_{R})+F(\pi,T_{L})-F(0,T_{L})=F_{T_{R}}+F_{T_{L}}$, where $F_{T_{R}}=F(0,T_{R})-F(\pi,T_{R})$ and $F_{T_{L}}=F(\pi,T_{L})-F(0,T_{L})$.
For spin flip scattering, the magnitude of Josephson current is much larger than for no flip case, see Fig.~6 where we plot Josephson current as function of phase difference $\varphi$ for both no flip and spin flip processes. Free energy is the integral of Josephson current. Thus, the magnitude of Free energy (see inset of Fig.~5(a)) and the Free energy difference $F_{T_{R}}$ and $F_{T_{L}}$ at specific temperatures are much larger for spin-flip process than no flip process.}
\begin{figure}[ht]
\centering{\includegraphics[width=.7\linewidth]{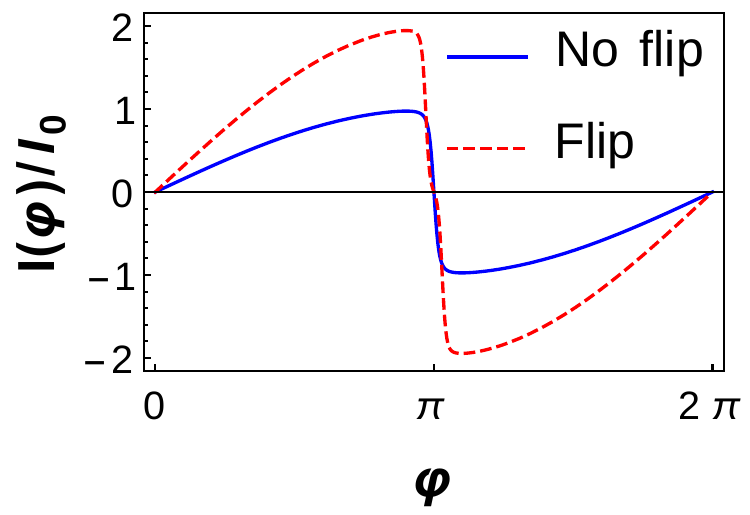}}
\caption{\small \sl {Josephson current as function of phase difference $\varphi$. Parameters: $J=0.1$, Flip: $S=-m'=1/2$, $p=1$; No Flip: $S=m'=1/2$, $p=0$, $k_{B}T=0.01\Delta$, $I_{0}=e\Delta/\hbar$.}}
\end{figure}
In presence of spin-flip scattering magnitude of $W$ at $\varphi_{f}=\pi$, i.e., {$W=0.735\Delta$} is maximum work done by our system. From inset of Fig.~5(a), we notice that Free energy is minimum at $\varphi=0$, i.e., when system shows $0$-junction behavior. From Fig.~5(c), we notice that at $\varphi_{f}=\pi$, efficiency $\eta$ for no flip case is a little bit larger than spin-flip case. Thus, spin flip process is better for work done, but no flip process is better for efficiency. 
{For our system total work done $W$ is larger for spin flip case as compared to no flip case. Since the magnitude of Josephson current and Free energy for spin flip case are larger which enhances the work done. However, efficiency is defined as $\eta=\frac{W}{Q_{R}}$, where $Q_{R}$ is the heat exchanged with right reservoir. We find that in presence of spin flip scattering $Q_{R}$ is larger for spin flip case than for no-flip case similar to Josephson current. However, the ratio $\frac{W}{Q_{R}}$ at $\varphi_{f}=\pi$ for no flip case is slightly larger than spin flip case. Thus, in this paper when $W$ is maximum, the corresponding efficiency $\eta$ is larger for no flip case than spin flip case.} In Figs.~5(b) and 5(d) we show dependence of work done and efficiency on $T_{R}$ for fixed $k_{B}T_{L}=0.01\Delta$ and compare the cycle's performance for no flip and spin flip scattering. In addition we plot Carnot efficiency in Fig.~5(d). Three important take home messages from Fig.~5 are (i) work done is maximum at $\varphi_{f}=\pi$ (see Fig.~5(a)), (ii) spin-flip scattering enhances work done (see Fig.~5(a)) and (iii) at {$k_{B}T_{R}=0.32\Delta$}, work done is maximum (see Fig.~5(b)). 

From Fig.~5(d), we notice that for small values of $T_{R}$, efficiency of a Josephson-Stirling engine is equal to Carnot efficiency $\eta_{C}$ regardless of spin-flip scattering, while at large values of $T_{R}$ in absence of spin-flip scattering, engine is more efficient and efficiency is close to Carnot limit. The reason we take low values of exchange coupling $J$ ($=0.1$) is because for these values of $J$ we get maximum work output \textquoteleft$W$\textquoteright{} and efficiency \textquoteleft$\eta$\textquoteright.
\begin{figure}[ht]
\centering{\includegraphics[width=.99\linewidth]{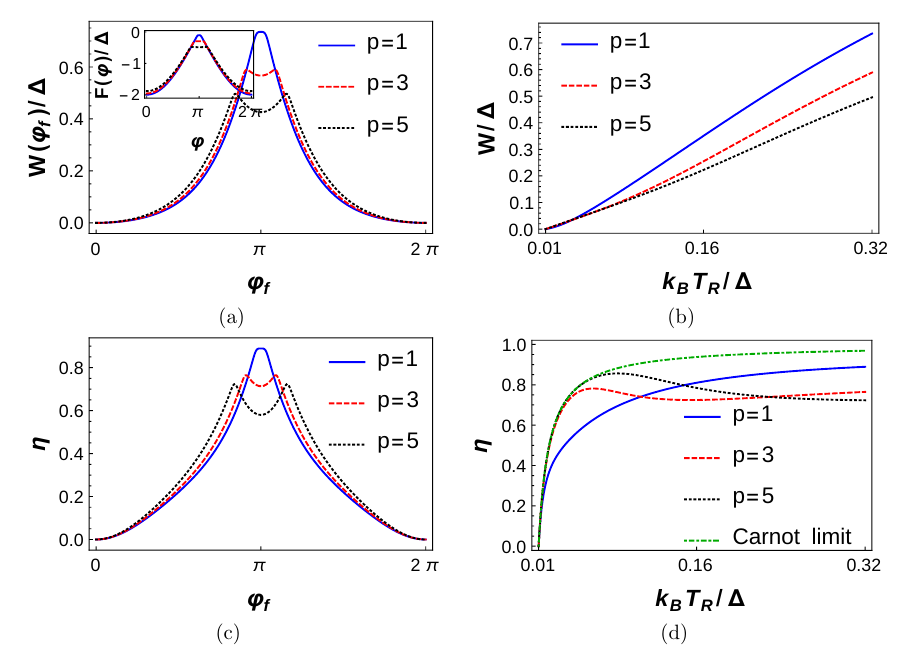}}
\caption{\small \sl  Total work done $W$ and efficiency $\eta$ of a Josephson-Stirling engine as function of $\varphi_{f}$ (for (a,c)) and as function of $T_{R}$ (for (b,d)) in presence of spin-flip scattering. Parameters: $J=0.1$,  $\varphi_{i}=0$, $\varphi_{f}=\pi$ (for $p=1$), $\varphi_{f}=0.91\pi$ (for $p=3$), $\varphi_{f}=0.84\pi$ (for $p=5$), $k_{B}T_{L}=0.01\Delta$, {$k_{B}T_{R}=0.32\Delta$} (for (a,c)), {$k_{B}T=0.01\Delta$.}}
\end{figure}
Next, in Fig.~7 we plot total work done $W$ and efficiency $\eta$ of a Josephson-Stirling engine for different levels of spin-flip scattering. In Figs.~7(a) and 7(c) we plot $W$ and $\eta$ as function of $\varphi_{f}$. We see that at $\varphi_{f}=\pi$ and for low values of spin-flip scattering, both work done and efficiency attain their maximum. However for high values of spin-flip scattering maximum values of work done and efficiency are no longer at $\varphi_f=\pi$. For flip probability $p=3$, $W$ and $\eta$ are maximum at two different values of $\varphi_f$,  $\varphi_f=0.91\pi$ and $\varphi_f=1.09\pi$, while for $p=5$, $W$ and $\eta$ are maximum at $\varphi_f=0.84\pi$ and $\varphi_f=1.16\pi$. The reason of this kind of behavior can be understood from Andreev bound state plots. For low values of spin-flip scattering two Andreev bound states merge near at $\varphi=\pi$ and thus $W$ is maximum at $\varphi_f=\pi$. {However for high values of spin-flip scattering maximum values of work done and efficiency are no longer at $\varphi_f=\pi$. For flip probability $p=3$, $W$ and $\eta$ are maximum at two different values of $\varphi_f$,  $\varphi_f=0.91\pi$ and $\varphi_f=1.09\pi$, while for $p=5$, $W$ and $\eta$ are maximum at $\varphi_f=0.84\pi$ and $\varphi_f=1.16\pi$. The reason of this kind of behavior can be understood from Andreev bound state plots. For no flip or low values of spin-flip scattering two Andreev bound states touch at $\varphi=\pi$ or near $\varphi=\pi$ and thus $W$ is maximum at $\varphi_f=\pi$. As spin-flip scattering breaks Andreev bound state energy degeneracy, therefore for high values of spin-flip scattering two Andreev bound states no longer touch at $\varphi=\pi$, but rather they touch at two locations symmetrically located around $\varphi=\pi$. For $p=3$, two Andreev bound state energies touch at $\varphi=0.91\pi$ and $\varphi=1.09\pi$, while for $p=5$ two Andreev bound state energies touch at $\varphi=0.84\pi$ and $\varphi=1.16\pi$. When two Andreev bound state energies touch, work done and efficiency attain their maximum values. Thus, with increase of flip probability of spin flipper, maximum values of $W$ and $\eta$ are no longer at $\varphi_f=\pi$ but are shifted with respect to it. Thus, with increase of flip probability of spin flipper, $W$ and $\eta$ are maximum at two distinct values of $\varphi_{f}$, therefore our device can be operated at both these values and not just at $\varphi=\pi$.} In inset of Fig.~7(a), we notice that minimum of Free energy occurs at $\varphi=0$, i.e., system shows $0$-junction behavior when it operates as a Josephson quantum heat engine. In Figs.~7(b) and 7(d) we plot $W$ and $\eta$ as function of $T_{R}$ for different levels of spin flip scattering at fixed $k_{B}T_{L}=0.01\Delta$. {The efficiency of our device is always either equal or less than Carnot efficiency as seen in Fig.~7(d).}
We choose $\varphi_{f}$ in such a way that $W$ becomes maximum. We see that for low values of spin-flip scattering, both work done and efficiency are larger. 
\subsubsection{Condition for optimality} 
Further we find in Fig.~7 that, at {$k_{B}T_{R}=0.32\Delta$}, both $W$ and $\eta$ are maximum. Thus, for spin-flip probability, $p=1$, $J=0.1$, $\varphi_{f}=\pi$, $k_{B}T_{L}=0.01\Delta$ and {$k_{B}T_{R}=0.32\Delta$} both work done and efficiency of our Josephson quantum heat engine are larger and therefore this regime is an optimal operating range of our Josephson quantum heat engine. {Please note that in Fig.~7 we consider the upper limit of reservoir temperatures is $3.7$K, which is much smaller than the superconducting critical temperature $T_{c}$ so that the superconducting gap $\Delta$ is not affected by temperature. Therefore, we can not make any comment what will happen at higher temperatures.}
This condition of optimality for our case is contrasted with that in Ref.~\cite{bsc}. In Ref.~\cite{bsc}, parity-conserving heat engine operates in an optimal operating point at $\varphi_{f}=2\pi$, while non-parity-conserving heat engine works in an optimal operating point at $\varphi_{f}=\pi$.

\subsection{Josephson-Stirling cycle as quantum refrigerator} 
Till now we have discussed work done and efficiency when Josephson-Stirling cycle acts as a quantum heat engine. In Fig.~8 we show action of Stirling cycle as a Josephson quantum refrigerator and calculate work done and coefficient of performance (COP). In Figs.~8(a) and 8(b), total work absorbed by refrigerator and COP are plotted as function of $\varphi_{f}$, for $T_{R}<T_{L}$ ({$k_{B}T_{L}=0.32\Delta$ and $k_{B}T_{R}=0.31\Delta$}). We see that for low values of spin flip scattering both $|W|$ and COP are larger and the product of COP and $|W|$ is maximum at $\varphi_{f}=\pi$. Thus, Josephson quantum refrigerator works in an optimal operating range for parameters of Fig.~8. For optimal parameters, {$|W|=0.02\Delta$} and {$\mbox{COP}=30.61$}. Although, our system exhibits maximum $|W|$ of {$0.62\Delta$} at {$k_{B}(T_{L}-T_{R})=0.25\Delta$}{, however maximum COP is same with the optimal value} as shown in Figs.~8(c) and 8(d) respectively.
\begin{figure}[ht]
\centering{\includegraphics[width=.99\linewidth]{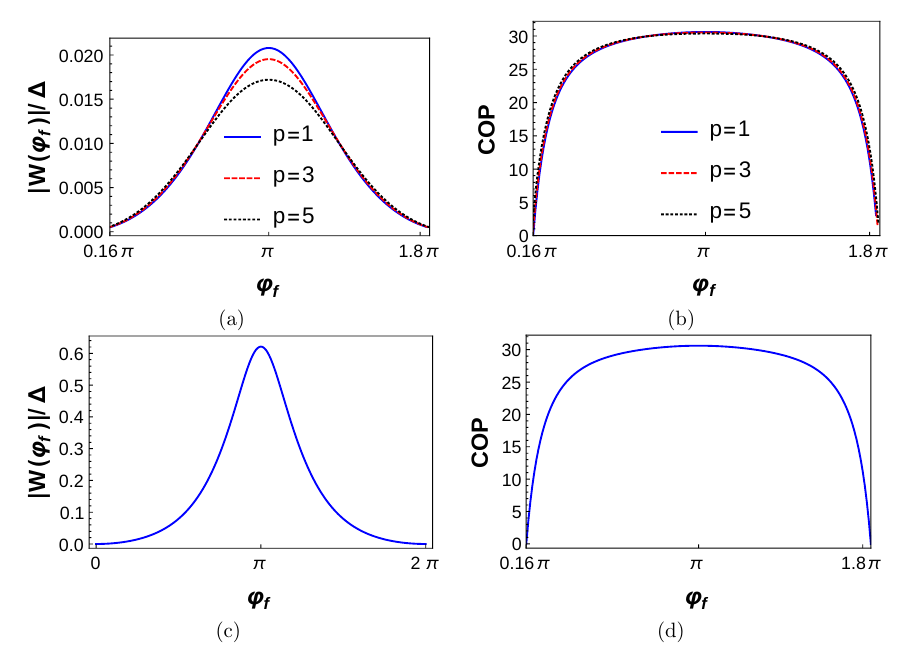}}
\caption{\small \sl (a,c) Total work $W$ absorbed, and (b,d) COP of quantum refrigerator as function of $\varphi_{f}$ in presence of spin-flip scattering. Parameters: $J=0.1$, {$k_{B}T_{L}=0.32\Delta$} (for (a,b)), {$k_{B}T_{R}=0.31\Delta$} (for (a,b)), $S=-m'=1/2$ (for (c,d)), $\varphi_{i}=0$.}
\end{figure}

\subsection{Phase diagram of Josephson-Stirling cycle} 
In Figs.~5,7,8 we depict the performance of this device as a quantum heat engine or quantum refrigerator depending on relative temperatures of reservoirs. In this subsection we will discuss phase diagram of a Josephson-Stirling cycle as function of phase difference $\varphi_{f}$ during the cycle. In Fig.~9, we plot $W$, $Q_{R}$, $Q_{L}$, $\eta$ and COP as functions of $\varphi_{f}$ and $T_{R}$ for low values of spin flip scattering ($S=1/2$, $m'=-1/2$, $p=1$) and high values of exchange coupling ($J=2$) of spin flipper. We fix $k_{B}T_{L}=0.01\Delta$ and consider $T_{R}>T_{L}$.
\begin{figure}[ht]
\centering{\includegraphics[width=.99\linewidth]{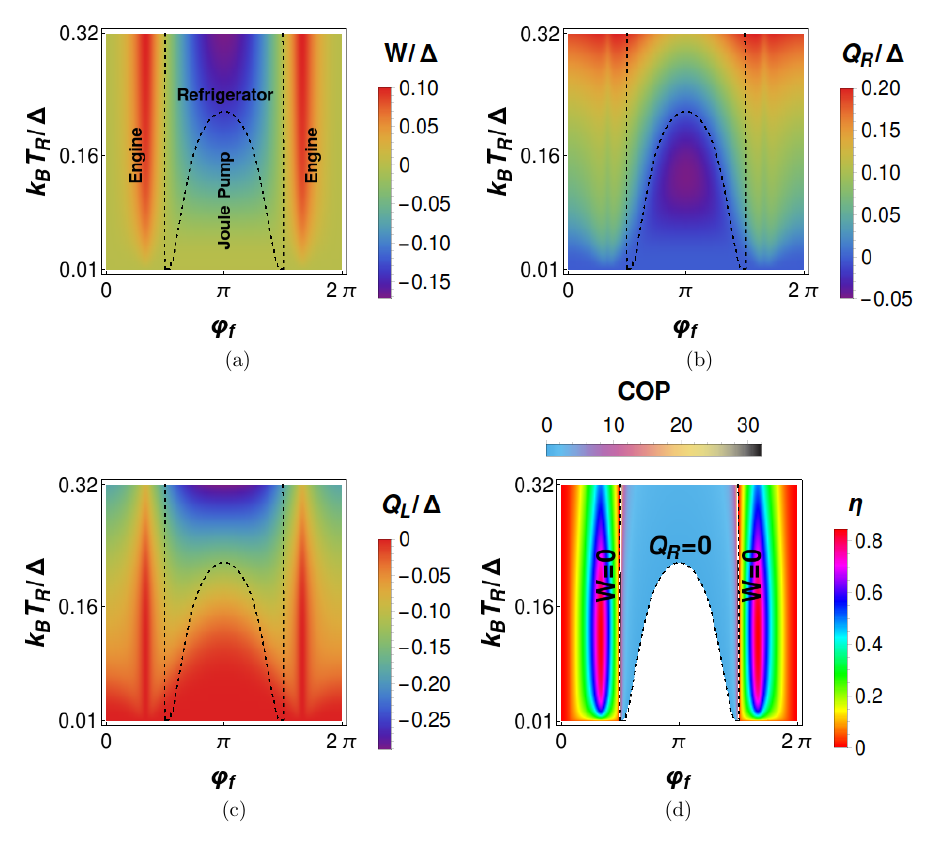}}
\caption{\small \sl (a) Total work $W$ (b) $Q_{R}$ and (c) $Q_{L}$, (d) $\eta$ or COP as function of $\varphi_{f}$ and $T_{R}$ in presence of spin flip scattering. Parameters: $S=1/2$, $m'=-1/2$, $J=2$, $k_{B}T_{L}=0.01\Delta$, $\varphi_{i}=0$.}
\end{figure}
From Fig.~9(a) we notice that with change of  $\varphi_{f}$ there is a sign change in $W$. $W$ changes sign from $W>0$ to $W<0$ between $\varphi_{f}=0$ and $\varphi_{f}=\pi$ and again changes sign from $W<0$ to $W>0$ between $\varphi_{f}=\pi$ and $\varphi_{f}=2\pi$. The regime where $W$ is positive ($W>0$) in Fig.~9(a), $Q_{R}$ is positive (Fig.~9(b)) but $Q_{L}$ is negative (Fig.~9(c)), in these regimes Josephson-Stirling cycle acts as an engine. Similarly in regime, wherein both $W$ and $Q_{L}$ are negative, cycle acts as a refrigerator if $Q_{R}$ is positive while cycle acts as a Joule pump if $Q_{R}$ is negative. A Joule pump transforms work into heat released to the reservoirs. Thus, there is a transition from engine mode to refrigerator or Joule pump mode with change of $\varphi_{f}$. Since, $\varphi_{f}$ is controlled by magnetic flux $\Phi$, therefore by changing the enclosed flux $\Phi$ in the Josephson junction loop we can tune our system from engine mode to refrigerator/Joule pump mode which is an attractive feature of our device. 
\subsection{Why is spin-flip scattering necessary?}
In this section we show why spin-flip scattering is necessary to tune our system from engine mode to refrigerator/Joule pump mode by the externally applied magnetic flux.

In absence of spin-flip scattering ($p=0$), we find work done (see Eq.~\eqref{worknoflip}) $W>0$ since, 
\small{
\begin{eqnarray}
(y_{L}\ln[\cosh(x_{L})]-y_{R}\ln[\cosh(x_{R})]) &>& (y_{L}\ln[\cosh(x_{L}E(\varphi_{f})/\Delta)]\nonumber\\
-y_{R}\ln[\cosh(x_{R}E(\varphi_{f})/\Delta)]),
\end{eqnarray}}
\normalsize for $T_{R}>T_{L}$ and regardless of $\varphi_{f}$. Thus, $W$ is always positive ($W>0$) in absence of spin-flip scattering irrespective of $\varphi_{f}$ and the Josephson-Stirling cycle operates solely as a quantum heat engine. {In absence of spin flip scattering Josephson current does not change sign between phase difference $\varphi=0$ and $\varphi=\pi$. Thus, for no flip process work done $W$ is always positive when $T_{R}>T_{L}$.} However, in presence of spin-flip scattering ($p\neq0$) and with $T_{R}>T_{L}$, we find $W$ (see Eq.~\eqref{work}) is not always positive and depending on other parameter values, $W$ can be negative ($W<0$) in certain range of $\varphi_{f}$. {In presence of spin flip scattering Josephson current can change sign between $\varphi=0$ and $\varphi=\pi$, see Fig.~10 where we plot Josephson current as function of phase difference $\varphi$ for both no flip and spin flip processes. {For low values of $J$, the system shows $0$-junction behavior irrespective of spin flip scattering (see Fig.~6), however for high values of $J$, the system shows $\pi$-junction behavior in presence of spin flip scattering as shown in Fig.~10. Thus, with increase of $J$ there is a transition from $0$ to $\pi$ junction in presence of spin flip scattering. In Fig.~11 we plot absolute value of Josephson current as function of exchange coupling $J$ for different values of flip probability $p$ of spin flipper. We see that in presence of spin flip scattering there is a discontinuous change as function of coupling $J$ in Josephson current due to $0$-$\pi$ junction transition, while no such change occurs in absence of spin flip scattering. For $p=1$, $0$-$\pi$ junction transition occurs at critical value $J=1.27$. Therefore, for $J=0.1$, in $0$ phase, Josephson current for no flip is less than spin flip case while at $J=2$ in $\pi$ phase Josephson current for no flip is more than that for flip.} When Josephson current changes sign, there will be a sign change in $W$ from positive ($W>0$) to negative ($W<0$) between $\varphi=0$ and $\varphi=\pi$. In the parameter regime wherein $W$ is negative and $Q_{L}$ is negative with $Q_{R}$ being positive, the Josephson-Stirling cycle operates as a refrigerator. Similarly, in parameter regime wherein $W$, $Q_{R}$ and $Q_{L}$ are all negative, Josephson-Stirling cycle acts as a Joule pump. Thus, in presence of spin-flip scattering when $W$ is negative, the Josephson-Stirling cycle can act as a refrigerator or Joule pump even when $T_{R}>T_{L}$. This is unique to this proposal since in other proposals\cite{bsc,carr}, $W$ is always positive and Josephson-Stirling cycle only acts as a heat engine when $T_{R}>T_{L}$.
\begin{figure}[ht]
\centering{\includegraphics[width=.65\linewidth]{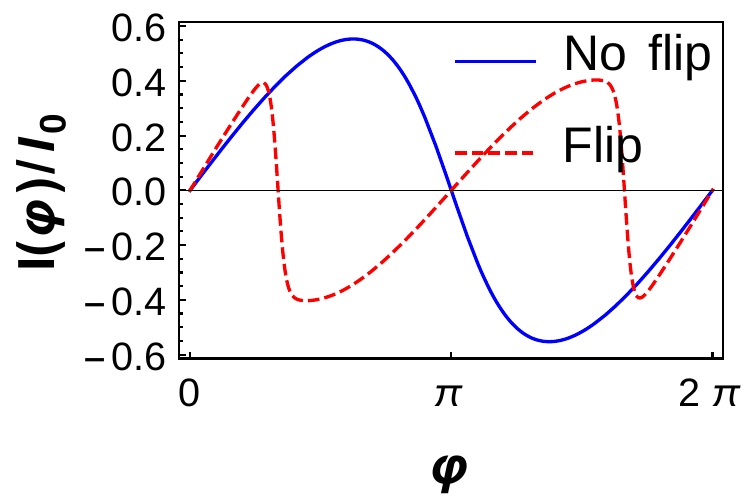}}
\caption{\small \sl {Josephson current as function of phase difference $\varphi$. Parameters: $J=2$, Flip: $S=-m'=1/2$, $p=1$; No flip: $S=m'=1/2$, $p=0$, $k_{B}T=0.01\Delta$, $I_{0}=e\Delta/\hbar$.}}
\end{figure}
\begin{figure}[ht]
\centering{\includegraphics[width=.65\linewidth]{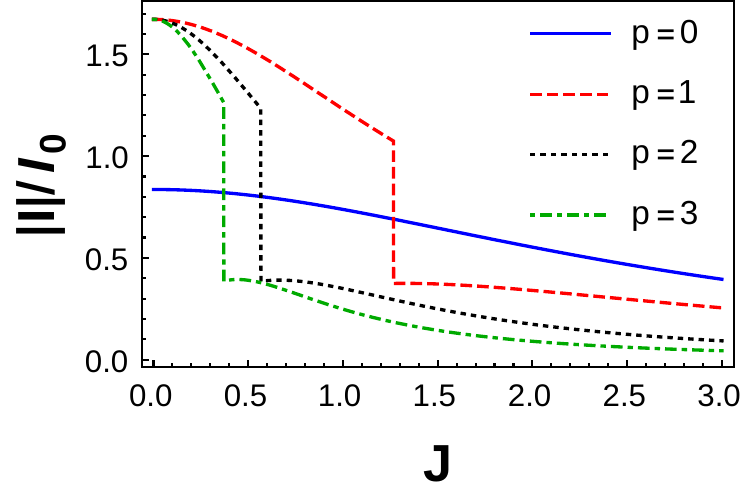}}
\caption{\small \sl {Absolute value of Josephson current as function of exchange coupling $J$ for different values of flip probability $p$ of spin flipper. Parameters: $k_{B}T=0.01\Delta$, $I_{0}=e\Delta/\hbar$.}}
\end{figure}
\begin{figure}[ht]
\centering{\includegraphics[width=.99\linewidth]{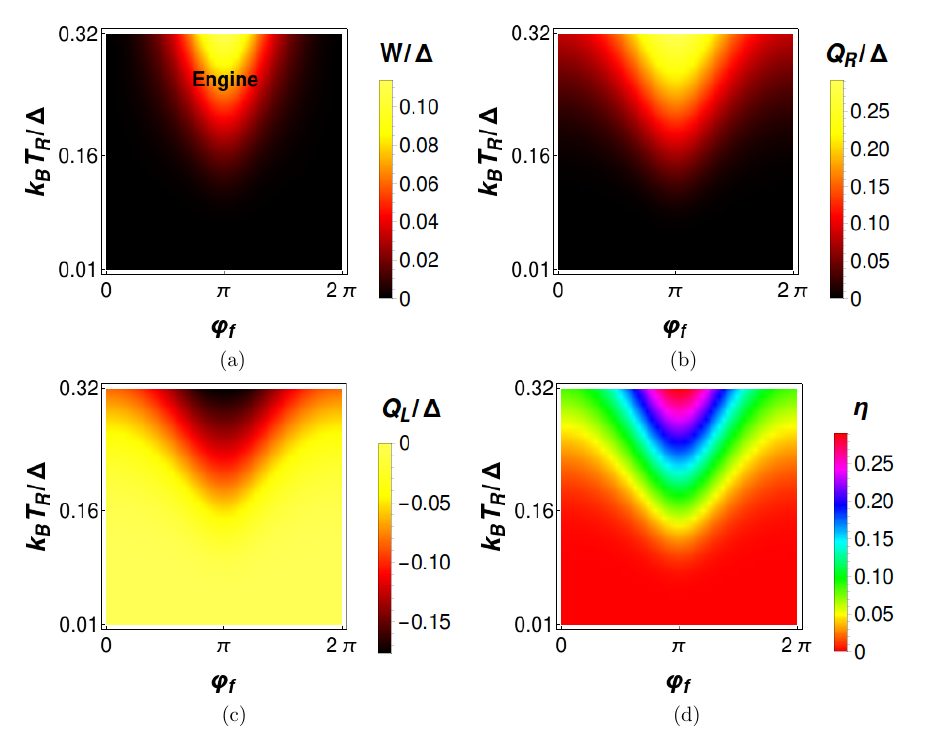}}
\caption{\small \sl (a) Total work done $W$ (b) $Q_{R}$ and (c) $Q_{L}$  (d) efficiency $\eta$ as function of maximal phase change $\varphi_{f}$ during Stirling cycle and right reservoir temperature $T_{R}$ in absence of spin-flip scattering. Other parameters are: $S=m'=1/2$, $J=2$, $k_{B}T_{L}=0.01\Delta$, $\varphi_{i}=0$.}
\label{spinflip-fig}
\end{figure}
In Fig.~\ref{spinflip-fig}, $W$, $Q_{R}$, $Q_{L}$ and $\eta$ are plotted as function of $\varphi_{f}$ and $T_{R}$ in absence of spin-flip scattering ($S=m'=1/2$, $p=0$). We see that $W$ does not change sign with $\varphi_{f}$ and thus there is no transition from engine mode to refrigerator or Joule pump mode. Therefore, we can conclude that it is spin flip scattering, which is responsible for tuning our system from engine mode to refrigerator/Joule pump mode via magnetic flux $\Phi$.
\begin{figure}[ht]
\centering{\includegraphics[width=.99\linewidth]{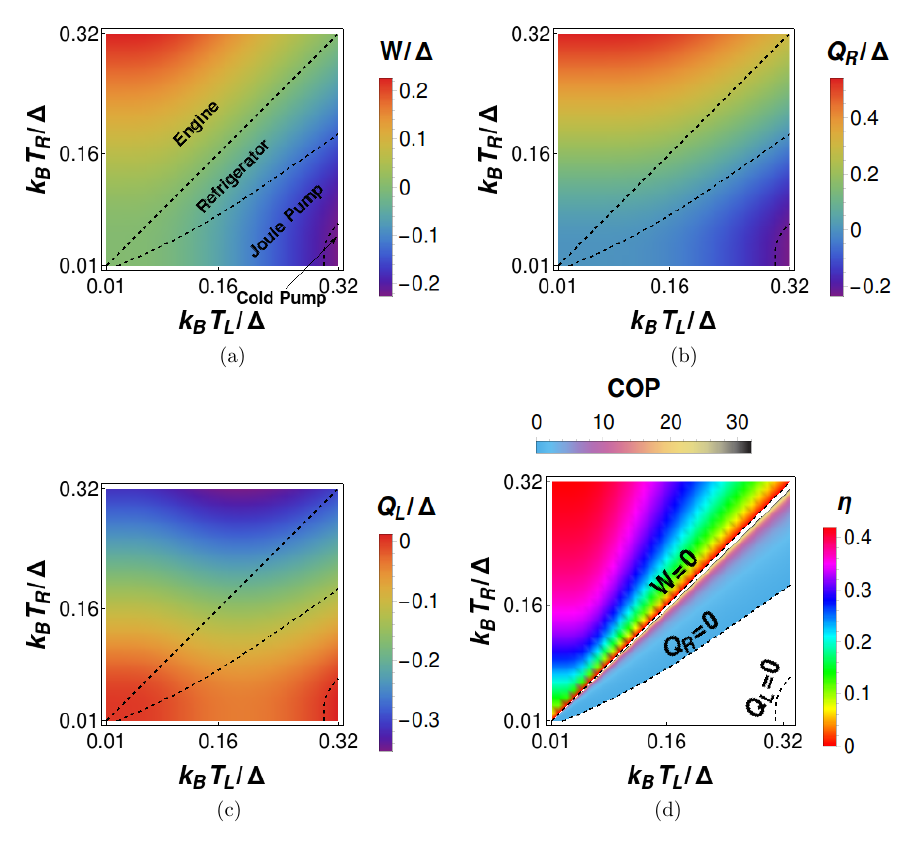}}
\caption{\small \sl (a) Total work done $W$, (b) heat exchanged $Q_{R}$ and (c) $Q_{L}$ and  (d) efficiency $\eta$ or COP as function of temperatures $T_{R}$ and $T_{L}$ in presence of spin-flip scattering. Parameters: $S=1/2$, $m'=-1/2$, {$J=1$}, $\varphi_{i}=0$, $\varphi_{f}=\pi$.}
\label{cold-pump}
\end{figure}
\subsection{Josephson-Stirling cycle as Joule pump and cold pump}
Finally, our device doesn't only exhibit the engine, refrigerator and Joule pump phases, it also exhibits a cold pump phase again in presence of spin-flip scattering only. The thin sliver in Fig.~\ref{cold-pump}(a) at bottom right corner shows the cold pump phase. In Fig.~\ref{cold-pump}, we plot $W$, $Q_{R}$, $Q_{L}$, efficiency of  engine and COP of refrigerator as functions of $T_{R}$ and $T_{L}$ for low values of spin flip scattering ($S=1/2$, $m'=-1/2$, $p=1$) and {high} values of exchange coupling ({$J=1$}) of spin flipper. We choose $\varphi_{f}$ such that $|W|$ will be maximized. We see that for $T_{R}>T_{L}$, the Josephson-Stirling cycle acts as an engine. {We take high values of $J$ ($J=1$) to see all phases in the phase diagram, however for low values of $J$ ($J=0.1$), at $k_{B}T_{L}=0.01\Delta$ and $k_{B}T_{R}=0.32\Delta$, the work done and efficiency in the engine mode attain their maximum values which are $W_{max}=0.735\Delta$ and $\eta_{max}=0.89$.} For $T_{R}<T_{L}$, we notice that the machine acts as a refrigerator or as Joule pump or cold pump depending on the sign of $Q_{R}$ and $Q_{L}$. The work done in the refrigerator mode is maximum at {$k_{B}T_{L}=0.32\Delta$} and {$k_{B}T_{R}=0.07\Delta$} and the maximum value is {$|W|_{max}=0.62\Delta$,} while the COP of the refrigerator is maximum at {$k_{B}T_{L}=0.32\Delta$} and {$k_{B}T_{R}=0.31\Delta$} and the maximum value is {$COP_{max}=30.61$}. Further, in the Joule pump mode the work done is maximum at {$k_{B}T_{L}=0.32\Delta$} and $k_{B}T_{R}=0.01\Delta$ and the maximum value is {$|W|_{max}=0.735\Delta$, while the maximum COP of the Joule pump is $COP_{max}=1$.} If $Q_{R}<0$, $W<0$ and $Q_{L}>0$, the cycle acts as a cold pump which transfers heat from the hot reservoir to the cold reservoir. The work done in the cold pump mode is maximum at {$k_{B}T_{L}=0.32\Delta$} and $k_{B}T_{R}=0.01\Delta$ and the maximum value is {$|W|_{max}=0.23\Delta$}, while the COP of the cold pump is maximum at {$k_{B}T_{L}=0.32\Delta$} and $k_{B}T_{R}=0.01\Delta$ and the maximum value is {$COP_{max}=1.05$}. Thus, by tuning the temperature of the reservoirs our system can be tuned from engine mode to refrigerator or Joule pump or cold pump mode.
\section{Analysis}
In this section we analyze in detail the expressions for work done (Eqs.~\eqref{work}, \eqref{worknoflip}) and heat exchanged with left and right reservoirs (Eqs.~\eqref{hrr}-\eqref{QLnoflip}). We also provide reasons why in absence of spin-flip scattering, one cannot tune the work done or the heat exchanged by the externally applied magnetic flux, showing the centrality of the spin-flipper to the device functioning.
\subsection{Work done ($W$)}
In presence of spin flip scattering, from Eq.~\eqref{work} we find that
\begin{widetext}
\begin{equation}
\label{Wflipcon}
\mbox{for work done $(W)>0$:}\,\,\,\,X^{\frac{T_{R}}{T_{L}}}>X^{\prime},\,\,\,\mbox{and}\,\,\,
\mbox{for $W<0$:}\,\,\,\,X^{\frac{T_{R}}{T_{L}}}<X^{\prime},
\end{equation}
where $X=\frac{e^{-\frac{F(\varphi_{f},T_{R})}{2k_{B}T_{R}}}}{D_{R}}$, $X^{\prime}=\frac{e^{-\frac{F(\varphi_{f},T_{L})}{2k_{B}T_{L}}}}{D_{L}}$, $D_{R}=4\cosh(x_{R})\cosh(x_{R}C^{\prime})$, and $D_{L}=4\cosh(x_{L})\cosh(x_{L}C^{\prime})$. From Eq.~\eqref{Wflipcon}, after some algebraic calculations we get,
\begin{align}
\begin{split}
\label{WFfre}
&\mbox{for $W>0$:}\,\,\,\,F(\varphi_{f},T_{L})-F(\varphi_{f},T_{R})>2k_{B}(T_{R}\ln[D_{R}]-T_{L}\ln[D_{L}]),\,\,\mbox{and}\\
&\mbox{for $W<0$:}\,\,\,\,F(\varphi_{f},T_{L})-F(\varphi_{f},T_{R})<2k_{B}(T_{R}\ln[D_{R}]-T_{L}\ln[D_{L}]).
\end{split}
\end{align}
\end{widetext}
{In our figures, the range of $2k_{B}(T_{R}\ln[D_{R}]-T_{L}\ln[D_{L}])$ for heat engine, refrigerator, Joule pump and cold pump modes are {$0$ to $0.36\Delta$, $-0.096\Delta$ to $0$, $-0.11\Delta$ to $0$, and $-0.11\Delta$ to $-0.10\Delta$} respectively, i.e., it always positive {for heat engine mode, while it always negative for refrigerator, Joule pump and cold pump modes.} Thus, when $W>0$, $F(\varphi_{f},T_{L})>F(\varphi_{f},T_{R})$ and when $W<0$, $F(\varphi_{f},T_{L})<F(\varphi_{f},T_{R})$. 
When work done is positive ($W>0$) the cycle operates as a heat engine, while when work done is negative ($W<0$) the cycle operates as a refrigerator or a Joule pump or a cold pump depending on the sign of $Q_{R}$ and $Q_{L}$ (see section III.A). Thus, Eqs.~\eqref{Wflipcon}, \eqref{WFfre} help us to understand the operating modes of our device. From Eqs.~\eqref{Wflipcon}, \eqref{WFfre}, we also see that by changing the reservoir temperatures $T_{R}$ and $T_{L}$ or phase $\varphi_{f}$ one can control the sign of $W$. Since $\varphi_{f}$ can be controlled via magnetic flux $\Phi$ enclosed in the Josephson junction loop, thus by controlling the temperature of reservoirs or magnetic flux $\Phi$, we can experimentally implement the condition for $W>0$ or $W<0$ in presence of spin flip scattering.

Herein, we prove that if spin flip scattering is absent, then there is only phase, the heat engine phase in our device, if temperatures $T_{R}$, $T_{L}$ are fixed and only flux changes. In absence of spin flip scattering from Eq.~\eqref{worknoflip} we get
\begin{widetext}
\begin{equation}
\label{WNCON}
\mbox{for $W>0$:}\,\,\,\,Y^{\frac{T_{R}}{T_{L}}}>Y^{\prime},\,\,\,\mbox{and}\,\,\,
\mbox{for $W<0$:}\,\,\,\,Y^{\frac{T_{R}}{T_{L}}}<Y^{\prime},
\end{equation}
%\end{widetext}
where $Y=\frac{e^{-\frac{F(\varphi_{f},T_{R})}{2k_{B}T_{R}}}}{2\cosh(x_{R})}$, and $Y^{\prime}=\frac{e^{-\frac{F(\varphi_{f},T_{L})}{2k_{B}T_{L}}}}{2\cosh(x_{L})}$. From Eq.~\eqref{WNCON}, after some algebraic calculations we get,
%\begin{widetext}
\begin{align}
\begin{split}
&\mbox{for $W>0$:}\,\,\,\,F(\varphi_{f},T_{L})-F(\varphi_{f},T_{R})>2k_{B}(T_{R}\ln[2\cosh(x_{R})]-T_{L}\ln[2\cosh(x_{L})]),\,\,\mbox{and}\\
&\mbox{for $W<0$:}\,\,\,\,F(\varphi_{f},T_{L})-F(\varphi_{f},T_{R})<2k_{B}(T_{R}\ln[\cosh(x_{R})]-T_{L}\ln[2\cosh(x_{L})]).
\end{split}
\end{align}
%\end{widetext}
In our figures, the range of $2k_{B}(T_{R}\ln[2\cosh(x_{R})]-T_{L}\ln[2\cosh(x_{L})])$ for engine mode is {$0-0.03\Delta$}. 
%\begin{widetext}
\begin{eqnarray}
\label{Wexp}
&&\mbox{If}\,\, G=Y^{\frac{T_{R}}{T_{L}}}-Y^{\prime}=\Bigg(\cosh\Bigg(\frac{x_{R}E(\varphi_{f})}{\Delta}\Bigg)\sech(x_{R})\Bigg)^{\frac{T_{R}}{T_{L}}}-\Bigg(\cosh\Bigg(\frac{x_{L}E(\varphi_{f})}{\Delta}\Bigg)\sech(x_{L})\Bigg),\,\, \mbox{then}\\ \nonumber\\ 
&&\frac{dG}{d\varphi_{f}}=\frac{x_{L}\sin(\varphi_{f})\Big(\sech(x_{L})\sinh\Big(\frac{x_{L}E(\varphi_{f})}{\Delta}\Big)-\Big(\cosh\Big(\frac{x_{R}E(\varphi_{f})}{\Delta}\Big)\sech(x_{R})\Big)^{\frac{T_{R}}{T_{L}}}\tanh\Big(\frac{x_{R}E(\varphi_{f})}{\Delta}\Big)\Big)}{\sqrt{4+J^2m'^2}\sqrt{2+J^2m'^2+2\cos(\varphi_{f})}}.
\label{Wderi}
\end{eqnarray}
\end{widetext}
When $G$ is maximum or minimum with respect to $\varphi_{f}$, then $\frac{dG}{d\varphi_{f}}=0$. Thus, from Eq.~\eqref{Wderi} we get $\sin(\varphi_{f})=0$ or $\varphi_{f}=0, \pi,\, \mbox{and}\,\, 2\pi$. For, $\varphi_{f}=0$ or $\varphi_{f}=2\pi$, we find that
\begin{widetext}
\begin{equation}
\label{Wdou}
\frac{d^2G}{d\varphi_{f}^2}=\frac{x_{L}(\tanh(x_{L})-\tanh(x_{R}))}{4+J^2m'^2}=\frac{\Delta\Big(\tanh\Big(\frac{\Delta}{2k_{B}T_{L}}\Big)-\tanh\Big(\frac{\Delta}{2k_{B}T_{R}}\Big)\Big)}{2k_{B}T_{L}(4+J^2m'^2)}. 
\end{equation}
\end{widetext}
In Eq.~\eqref{Wdou}, when $T_{R}>T_{L}$, $\tanh\Big(\frac{\Delta}{2k_{B}T_{L}}\Big)>\tanh\Big(\frac{\Delta}{2k_{B}T_{R}}\Big)$, thus $\frac{d^2G}{d\varphi_{f}^2}>0$ at $\varphi_{f}=0$ or $\varphi_{f}=2\pi$. Therefore, at $\varphi_{f}=0$ or $\varphi_{f}=2\pi$, $G$ is minimum and from Eq.~\eqref{Wexp} the minimum value of $G$ is $G_{min}=0$. Since, the minimum value of $G$ is zero for $T_{R}>T_{L}$, thus $G$ is positive or $Y^{\frac{T_{R}}{T_{L}}}>Y^{\prime}$ irrespective of $\varphi_{f}$ for $T_{R}>T_{L}$. Thus, we can conclude that for no flip process, although the magnitude of $W$ depends on both reservoir temperatures and phase $\varphi_{f}$, but the sign of $W$ depends only on temperature of the reservoirs in contrast to spin flip case where the sign of $W$ depends on both reservoir temperatures and $\varphi_{f}$.
Similarly, for $T_{R}<T_{L}$, $\frac{d^2G}{d\varphi_{f}^2}<0$ at $\varphi_{f}=0$ or $\varphi_{f}=2\pi$. Thus, at $\varphi_{f}=0$ or $\varphi_{f}=2\pi$, $G$ is maximum when $T_{R}<T_{L}$ and from Eq.~\eqref{Wexp} the maximum value of $G$ is $G_{max}=0$.
Since, the maximum value of $G$ is zero for $T_{R}<T_{L}$, thus $G$ is negative or $Y^{\frac{T_{R}}{T_{L}}}<Y^{\prime}$ irrespective of $\varphi_{f}$ for $T_{R}<T_{L}$. In absence of spin flip scattering, since the sign of $W$ can not be tuned via magnetic flux, thus our device can not be tuned from heat engine mode to other operating mode like refrigerator or Joule pump or cold pump by changing magnetic flux.
\subsection{Heat exchanged (with right reservoir)}
Similarly, in presence of spin flip scattering from Eq.~\eqref{hrr} we find that
\begin{widetext}
\begin{align}
\begin{split}
\label{QRflipcon}
&\mbox{for $Q_{R}>0$:}\,\,\,Xe^{\frac{F(\varphi_{f},T_{R})}{2k_{B}T_{R}}+\frac{\Omega(\varphi_{f},T_{R})}{2k_{B}}+x_{R}C^{\prime}\tanh(x_{L}C^{\prime})+x_{R}\tanh(x_{L})}>1,\,\,\mbox{and}\\
&\mbox{for $Q_{R}<0$:}\,\,\,Xe^{\frac{F(\varphi_{f},T_{R})}{2k_{B}T_{R}}+\frac{\Omega(\varphi_{f},T_{R})}{2k_{B}}+x_{R}C^{\prime}\tanh(x_{L}C^{\prime})+x_{R}\tanh(x_{L})}<1,
\end{split}
\end{align}
From Eq.~\eqref{QRflipcon}, after some algebraic calculations we get,
\begin{align}
\begin{split}
\label{QRFLIPCON}
&\mbox{for $Q_{R}>0$:}\,\,\,\Omega(\varphi_{f},T_{R})>2k_{B}(\ln[D_{R}]-G_{R}) ,\,\,\mbox{and}\\
&\mbox{for $Q_{R}<0$:}\,\,\,\Omega(\varphi_{f},T_{R})<2k_{B}(\ln[D_{R}]-G_{R}),
\end{split}
\end{align}
\end{widetext}
where $G_{R}=x_{R}C^{\prime}\tanh(x_{L}C^{\prime})+x_{R}\tanh(x_{L})$. In our figures, the range of $2k_{B}(\ln[D_{R}]-G_{R})$ for heat engine, refrigerator, Joule pump and cold pump modes are {$0-1.49k_{B}$, $0-1.34k_{B}$, $0-20.62k_{B}$, and $3.97k_{B}-23.84k_{B}$} respectively, i.e., it always positive. Thus, when $Q_{R}>0$, $\Omega(\varphi_{f},T_{R})>0$. In Eqs.~\eqref{QRflipcon}, \eqref{QRFLIPCON}, the sign of $Q_{R}$ depends on both reservoir temperatures $T_{R}$, $T_{L}$ and phase $\varphi_{f}$. When $Q_{R}$ is positive ($Q_{R}>0$) the cycle operates as a heat engine or refrigerator depending on the sign of $W$, while when $Q_{R}$ is negative ($Q_{R}<0$) the cycle operates as a Joule pump or cold pump depending on the sign of $Q_{L}$ (see section III.A). Thus, the sign of $Q_{R}$ helps to understand the different operating modes of our device.

In absence of spin flip scattering from Eq.~\eqref{QRnoflip} we get,
\begin{widetext}
\begin{align}
\begin{split}
\label{QRnocon}
&\mbox{for $Q_{R}>0$:}\,\,\,Ye^{\frac{F(\varphi_{f},T_{R})}{2k_{B}T_{R}}+\frac{\Omega(\varphi_{f},T_{R})}{2k_{B}}+x_{R}\tanh(x_{L})}>1,\,\,\mbox{and}\\
&\mbox{for $Q_{R}<0$:}\,\,\,Ye^{\frac{F(\varphi_{f},T_{R})}{2k_{B}T_{R}}+\frac{\Omega(\varphi_{f},T_{R})}{2k_{B}}+x_{R}\tanh(x_{L})}<1.
\end{split}
\end{align}
\end{widetext}
From Eq.~\eqref{QRnocon}, after some algebraic calculations we get,
\begin{widetext}
\begin{align}
\begin{split}
&\mbox{for $Q_{R}>0$:}\,\,\,\Omega(\varphi_{f},T_{R})>2k_{B}(\ln[2\cosh(x_{R})]-x_{R}\tanh(x_{L})) ,\,\,\mbox{and}\\
&\mbox{for $Q_{R}<0$:}\,\,\,\Omega(\varphi_{f},T_{R})<2k_{B}(\ln[2\cosh(x_{R})]-x_{R}\tanh(x_{L})).
\end{split}
\end{align}
\end{widetext}
In our figures, the range of $2k_{B}(\ln[2\cosh(x_{R})]-x_{R}\tanh(x_{L}))$ for heat engine mode is {$0-0.32k_{B}$}. Since, in no flip process there is no transition from heat engine mode to other operating mode like refrigerator or Joule pump or cold pump with change of $\varphi_{f}$, thus the sign of $Q_{R}$ does not change with $\varphi_{f}$.
\subsection{Heat exchanged (with left reservoir)}
Finally, in presence of spin flip scattering from Eq.~\eqref{hlr} we get, 
\begin{widetext}
\begin{align}
\begin{split}
\label{QLflipcon}
&\mbox{for $Q_{L}>0$:}\,\,\,X^{\prime}e^{\frac{F(\varphi_{f},T_{R})}{2k_{B}T_{L}}+\frac{\Omega(\varphi_{f},T_{R})T_{R}}{2k_{B}T_{L}}+x_{L}C^{\prime}\tanh(x_{L}C^{\prime})+x_{L}\tanh(x_{L})}<1,\,\,\mbox{and}\\
&\mbox{for $Q_{L}<0$:}\,\,\,X^{\prime}e^{\frac{F(\varphi_{f},T_{R})}{2k_{B}T_{L}}+\frac{\Omega(\varphi_{f},T_{R})T_{R}}{2k_{B}T_{L}}+x_{L}C^{\prime}\tanh(x_{L}C^{\prime})+x_{L}\tanh(x_{L})}>1,
\end{split}
\end{align}
%\end{widetext}
From Eq.~\eqref{QLflipcon}, after some algebraic calculations we get,
%\begin{widetext}
\begin{align}
\begin{split}
\label{QLFLIPCON}
&\mbox{for $Q_{L}>0$:}\,\,\,(F(\varphi_{f},T_{R})-F(\varphi_{f},T_{L}))+\Omega(\varphi_{f},T_{R})T_{R}<2k_{B}T_{L}(\ln[D_{L}]-G_{L}) ,\,\,\mbox{and}\\
&\mbox{for $Q_{L}<0$:}\,\,\,(F(\varphi_{f},T_{R})-F(\varphi_{f},T_{L}))+\Omega(\varphi_{f},T_{R})T_{R}>2k_{B}T_{L}(\ln[D_{L}]-G_{L}),
\end{split}
\end{align}
\end{widetext}
where $G_{L}=x_{L}C^{\prime}\tanh(x_{L}C^{\prime})+x_{L}\tanh(x_{L})$. In our work, the range of $2k_{B}T_{L}(\ln[D_{L}]-G_{L})$ for heat engine, refrigerator, Joule pump and cold pump modes are {$0-3.87k_{B}$, $0-4.05k_{B}$, $0-4.05k_{B}$, and $3.74k_{B}-4.05k_{B}$} respectively, i.e., it always positive. Thus, when $Q_{L}<0$, $(F(\varphi_{f},T_{L})-F(\varphi_{f},T_{R}))<\Omega(\varphi_{f},T_{R})T_{R}$. In Eqs.~\eqref{QLflipcon}, \eqref{QLFLIPCON}, the sign of $Q_{L}$ depends on both reservoir temperatures $T_{R}$, $T_{L}$ and phase $\varphi_{f}$. When $Q_{L}$ is positive ($Q_{L}>0$) the cycle operates as a cold pump, while when $Q_{L}$ is negative ($Q_{L}<0$) the cycle operates as a heat engine or refrigerator or Joule pump depending on the sign of $W$ and $Q_{R}$ (see section III.A). Thus, similar to $Q_{R}$, the sign of $Q_{L}$ also helps to understand the different operating modes of our device. 

In absence of spin flip scattering from Eq.~\eqref{QLnoflip} we find that
\begin{widetext}
\begin{align}
\begin{split}
\label{QLnocon}
&\mbox{for $Q_{L}>0$:}\,\,\,Y^{\prime}e^{\frac{F(\varphi_{f},T_{R})}{2k_{B}T_{L}}+\frac{\Omega(\varphi_{f},T_{R})T_{R}}{2k_{B}T_{L}}+x_{L}\tanh(x_{L})}<1,\,\,\mbox{and}\\
&\mbox{for $Q_{L}<0$:}\,\,\,Y^{\prime}e^{\frac{F(\varphi_{f},T_{R})}{2k_{B}T_{L}}+\frac{\Omega(\varphi_{f},T_{R})T_{R}}{2k_{B}T_{L}}+x_{L}\tanh(x_{L})}>1,
\end{split}
\end{align}
%\end{widetext}
From Eq.~\eqref{QLnocon}, after some algebraic calculations we get,
%\begin{widetext}
\begin{align}
\begin{split}
&\mbox{for $Q_{L}>0$:}\,\,\,(F(\varphi_{f},T_{R})-F(\varphi_{f},T_{L}))+\Omega(\varphi_{f},T_{R})T_{R}<2k_{B}T_{L}(\ln[2\cosh(x_{L})]-x_{L}\tanh(x_{L})) ,\,\,\mbox{and}\\
&\mbox{for $Q_{L}<0$:}\,\,\,(F(\varphi_{f},T_{R})-F(\varphi_{f},T_{L}))+\Omega(\varphi_{f},T_{R})T_{R}>2k_{B}T_{L}(\ln[2\cosh(x_{L})]-x_{L}\tanh(x_{L})).
\end{split}
\end{align}
\end{widetext}
In our work, the range of $2k_{B}T_{L}(\ln[2\cosh(x_{L})]-x_{L}\tanh(x_{L}))$ for heat engine mode is {$0-1.17k_{B}$}. Since, in no flip process there is no transition between different operating modes of the Josephson-Stirling cycle with change of $\varphi_{f}$, thus the sign of $Q_{L}$ does not change with $\varphi_{f}$.
\section{Experimental realization and Conclusions}
Experimentally, high spin molecules, for example, Fe$_{19}$-complex with a spin of $S=33/2$ can to a certain extent be a model for the spin flipper. It is to be noted that the internal dynamics of such a high spin molecule may be quite different from the spin-flipper considered here. Even then, the spin flipper can mimic the half-integer spin states ($S$) up to any arbitrary high value and the associated spin magnetic moment of the high spin molecule and the consequence of an electron interacting with such, to a large extent.
Our proposed Josephson quantum spin thermodynamic device acts as a heat engine or refrigerator or Joule pump or as a cold pump via spin-flip scattering and is experimentally realizable. Doping a magnetic adatom or spin-flipper in an one dimensional superconducting loop shouldn't be difficult, especially with a $s$-wave superconductor like Lead or Aluminum it should be perfectly possible. {Futher, thermal valves have been experimentally realized in a different setup using quantum point contacts produced on top of two dimensional (2D) electron gas\cite{amado}. In our work we can also apply a similar method to realize thermal valves.} Our proposed experimental scheme is suitable to measure the work done and heat exchanged in an individual realization of the cycle. Work done per cycle during each isothermal process can be experimentally determined via the current-phase relation. For a Josephson Stirling cycle work done during each isothermal process is given as $W_{if}=-\frac{\hbar}{2e}\int_{\varphi_{i}}^{\varphi_{f}}I(\varphi,T)d\varphi$, where $\varphi_{i}$ and $\varphi_{f}$ are the initial and final phases respectively. In isophasic process since phase difference $\varphi$ is constant, there is no work done in this process. Thus, to measure the work done experimentally one needs to control phase difference $\varphi$ across the junction, regulated by flux $\Phi$, as well as to know about the Josephson current $I$ flowing through the loop. This can be done using a scanning superconducting quantum interference device (SQUID) microscope to perform the measurements of current-phase relation of the junction\cite{bsc,iso,sha}. By applying a current through the field coil of SQUID sensor, the magnetic flux through Josephson junction loop can be tuned. This magnetic flux, controls the phase difference across JJ and induces a supercurrent in Josephson junction loop. The supercurrent leads to a signal which is measured by pickup loop of SQUID sensor\cite{iso,iko}.

We have compared our proposal with other Josephson quantum heat engines and refrigerators in Table I. While, Refs.~\onlinecite{bsc}, \onlinecite{carr} are very important works which laid down the principle of Josephson quantum heat engine and refrigerator, efficiency and coefficient of performance (COP) of these can none-the-less be still enhanced, and the tunability of the device improved as we show in our work.
\begin{table}[ht]
\scriptsize
\caption{Josephson junction based quantum heat engines and refrigerators} 
\begin{tabular}{|p{1.7cm}|p{1.7cm}|p{1.67cm}|p{1.38cm}|p{1.68cm}|}
\hline
& \multicolumn{2}{|c|}{Heat engine mode} & \multicolumn{2}{|c|}{Refrigerator mode} \\
\hline
& $W_{max}$ & $\eta_{max}$ & $|W_{max}|$ & $\mbox{COP}_{max}$ \\
\hline
The JJ device (Fig.~1) & {$0.735\Delta$} ($\Delta=1meV$) (Fig.~5(a)) & {$0.94$} (Fig.~5(c)) & {$0.62\Delta$} (Fig.~8(c))  & {$30.61$} (Fig.~8(d)) \\
\hline
Topological Josephson heat engine (Ref.~\onlinecite{bsc}) & $2\Delta$ ($\Delta=150\mu eV$) (Fig.~4(c) of Ref.~\onlinecite{bsc}) & $0.8$ (Figs.~4(b,d) of Ref.~\onlinecite{bsc}) & \hspace{0.5cm}$-$ & $20$  (Figs.~4(b,d) of Ref.~\onlinecite{bsc}) \\
\hline
Josephson heat engine (Ref.~\onlinecite{carr}) & $0.38\Delta$ ($\Delta=180eV$) (Fig.~14(a) of Ref.~\onlinecite{carr})  & $0.5$ (Fig.~15(a) of Ref.~\onlinecite{carr}) & \hspace{0.5cm}$-$ & $10$ (Fig.~15(a) of Ref.~\onlinecite{carr}) \\
\hline
\end{tabular}
\end{table}
In both Refs.~\onlinecite{bsc} and \onlinecite{carr}, the Josephson quantum heat engine and refrigerator can be tuned from engine mode to refrigerator mode via exchanging the temperatures of left and right reservoirs only, while in this proposal it is seen that heat engine to refrigeration mode transition or for that matter to Joule or cold pump can be effected by either the magnetic flux enclosed or by tuning the temperature of reservoirs. This makes spin flipper doped Josephson junction loop much more versatile as regards possible applications. As a quantum heat engine, the proposed device is much more efficient than previously proposed Josephson quantum heat engines\cite{bsc,carr}. Further, when operating as a quantum refrigerator, COP of the proposed device is higher than that seen in Refs.~\cite{bsc,carr}. In Table I, we see that although work done by Josephson heat engine proposed in Ref.~\onlinecite{carr} is larger than ours, but efficiency of the Josephson quantum heat engine discussed in this work is much larger than those of Josephson heat engines of Refs.~\onlinecite{bsc,carr}. Further, as a Josephson quantum refrigerator, COP of the proposed device is huge compared to other Josephson quantum refrigerator proposals.
To conclude, in this work we have shown that a 1D Josephson junction loop doped with a spin flipper, and attached to two thermal reservoirs at inequivalent temperatures via thermal valves can act as a Josephson quantum heat engine or  quantum refrigerator or as a Joule pump or even as a cold pump. {The main advantage of spin flipper in our device is that in presence of spin flip scattering we can tune our device} from engine mode to other operating modes like refrigerator or Joule pump/Cold pump not only via changing the temperatures of the reservoirs, but also via the enclosed flux in Josephson junction loop which is the most lucrative aspect of work, since this fact alone implies a much greater possibility of our proposal being experimentally realized. {Thus, spin flip scattering is necessary to enhance the tunability of our device.}

 \acknowledgments 
 This work was supported by the grants: 1. Josephson junctions with strained Dirac materials and their application in quantum information processing, SERB Grant No. CRG/20l9/006258, and 2. Nash equilibrium versus Pareto optimality in N-Player games, SERB MATRICS Grant No. MTR/2018/000070.
\appendix
\section{Action of spin-flipper}
Action of exchange operator $\vec{s}.\vec{S}$ on spin up electron-like quasi-particle spinor is shown below,
{
\footnotesize{
\begin{equation}
\label{eu}
\vec s.\vec S\begin{pmatrix}
               u\\
               0\\
               0\\
               v
              \end{pmatrix}\phi_{m'}^{S}=s_{z}S_{z}\begin{pmatrix}
               u\\
               0\\
               0\\
               v
              \end{pmatrix}\phi_{m'}^{S}+\frac{1}{2}s^{-}S^{+}\begin{pmatrix}
               u\\
               0\\
               0\\
               v
              \end{pmatrix}\phi_{m'}^{S}+\frac{1}{2}s^{+}S^{-}\begin{pmatrix}
               u\\
               0\\
               0\\
               v
              \end{pmatrix}\phi_{m'}^{S}. 
\end{equation}}
\normalsize $s^{+}\begin{pmatrix}
               u\\
               0\\
               0\\
               v
              \end{pmatrix}=0$, as $s^{+}$ is the spin raising operator for electronlike quasiparticle and there are no higher spin states for a spin-$\frac{1}{2}$ electronlike quasiparticle than up and thus the 3rd term in Eq.~\eqref{eu} becomes zero, but $s^{-}\begin{pmatrix}
               u\\
               0\\
               0\\
               v
              \end{pmatrix}=\hbar\begin{pmatrix}
              0\\
              u\\
              -v\\
              0
             \end{pmatrix}$, the spin lowering operator gives the spin down electronlike quasiparticle state $\begin{pmatrix}
              0\\
              u\\
              -v\\
              0
             \end{pmatrix}$. Further, $s_{z}\begin{pmatrix}
               u\\
               0\\
               0\\
               v
              \end{pmatrix}=\frac{\hbar}{2}\begin{pmatrix}
               u\\
               0\\
               0\\
               v
              \end{pmatrix}$ for spin up electronlike quasiparticle and $S_{z}\phi_{m'}^{S}=\hbar m'\phi_{m'}^{S}$ for spin flipper. The spin-raising and spin-lowering operators operating on spin flipper give: $S^{+}\phi_{m'}^{S}=\hbar\sqrt{(S-m')(S+m'+1)}\phi_{m'+1}^{S}=\hbar p\phi_{m'+1}^{S}$ and $S^{-}\phi_{m'+1}^{S}=\hbar\sqrt{(S-m')(S+m'+1)}\phi_{m'}^{S}=\hbar p\phi_{m'}^{S}$.}
\small{              
\begin{equation}
\label{euu}
\mbox{ Thus, }\,\,\vec s.\vec S\begin{pmatrix}
               u\\
               0\\
               0\\
               v
              \end{pmatrix}\phi_{m'}^{S}=\frac{\hbar^2 m'}{2}\begin{pmatrix}
              u\\
              0\\
              0\\
              v
             \end{pmatrix}\phi_{m'}^{S}+\frac{\hbar^2 p}{2}\begin{pmatrix}
             0\\
             u\\
             -v\\
             0
             \end{pmatrix}\phi_{m'+1}^{S},
\end{equation}}
\normalsize where, $p=\sqrt{(S-m')(S+m'+1)}$ denotes flip probability\cite{AJP} for spin-flipper. Similarly, action of $\vec{s}.\vec{S}$ on the spin down electron-like quasi-particle spinor is
\small{
\begin{equation}
\label{ed}
\vec s.\vec S\begin{pmatrix}
               0\\
               u\\
               -v\\
               0
              \end{pmatrix}\phi_{m'+1}^{S}=-\frac{\hbar^2(m'+1)}{2}\begin{pmatrix}
              0\\
              u\\
              -v\\
              0
             \end{pmatrix}\phi_{m'+1}^{S}+\frac{\hbar^2 p}{2}\begin{pmatrix}
             u\\
             0\\
             0\\
             v
             \end{pmatrix}\phi_{m'}^{S}.
\end{equation}}
\normalsize Further, action of exchange operator $\vec{s}.\vec{S}$ on spin-up hole-like quasi-particle spinor is
\small{
\begin{equation}
\label{hu}
\vec s.\vec S\begin{pmatrix}
               0\\
               -v\\
               u\\
               0
              \end{pmatrix}\phi_{m'+1}^{S}=-\frac{\hbar^2 (m'+1)}{2}\begin{pmatrix}
              0\\
              -v\\
              u\\
              0
             \end{pmatrix}\phi_{m'+1}^{S}+\frac{\hbar^2 p}{2}\begin{pmatrix}
             v\\
             0\\
             0\\
             u
             \end{pmatrix}\phi_{m'}^{S},
\end{equation}}
\normalsize action of exchange operator $\vec{s}.\vec{S}$ on spin down hole-like quasi-particle spinor is
\small{
\begin{equation}
\label{hd}
\vec s.\vec S\begin{pmatrix}
               v\\
               0\\
               0\\
               u
              \end{pmatrix}\phi_{m'}^{S}=\frac{\hbar^2 m'}{2}\begin{pmatrix}
              v\\
              0\\
              0\\
              u
             \end{pmatrix}\phi_{m'}^{S}+\frac{\hbar^2 p}{2}\begin{pmatrix}
             0\\
             -v\\
             u\\
             0
             \end{pmatrix}\phi_{m'+1}^{S}.
\end{equation}}
\normalsize {In quantum spin flip scattering process, wherein Josephson supercurrent (state of Josephson supercurrent is given as $|s.c\rangle$) is denoted by a macroscopic wavefunction $\sim|\Psi_{S_{N}}|e^{i\varphi_{N}}\approx \begin{pmatrix}u\\
                              0\\
                              0\\ 
                              v
                              \end{pmatrix}e^{i\varphi_{N}}$ (where $N$ can be $L$ or $R$) interacts with the spin flipper, the spin flipper can flip its spin with finite probability, but there is no certainty for flipping its spin. In addition to the spin flip process, there is the other process of no flip. Thus, while before interaction, the supercurrent wavefunction and spin-flipper wavefunction are completely independent after interaction both are in a entangled and in a superposed state of:
{\begin{equation}
\label{ent}
\underbrace{\ket{s.c}\otimes\ket{\phi_{m'}^{S}}}_\textrm{Before interaction}\xrightarrow{\textrm{interaction}}\underbrace{{\frac{m'}{2}}\overbrace{\ket{\mbox{No-flip}}}^\textrm{Product state}+{\frac{p}{2}}\overbrace{\ket{\mbox{Mutual-flip}}}^\textrm{Entangled state}}_\textrm{After interaction}.
\end{equation}}
{From Eq.~\eqref{ent} we see that when Josephson supercurrent interacts with the spin flipper there is either a mutual spin-flip process in which an entangled state is formed or a no flip process in which a product state is formed. We will now explain separately how an entangled state will form in the mutual spin-flip process and how a product state will form in no flip process.}
The interaction of unpolarized Josephson current with spin-flipper leads to with finite flip probability an entangled state of spin up Josephson current and spin down spin-flipper \& spin down Josephson current and spin up spin-flipper and may also lead to with a product state of spin down Josephson current and spin down spin-flipper for no flip case,
\begin{widetext}
\begin{eqnarray}
&&\mbox{Spin flip:}\,\,\,\,\,\, \overbrace{\ket{\uparrow}_{s.c}}^\textrm{Josephson current state}\otimes\overbrace{\ket{\downarrow}_{\phi_{m'}^{S}}}^\textrm{Spin-flipper state}\xrightarrow{\textrm{interaction}}\overbrace{\ket{\downarrow}_{s.c}\otimes\ket{\uparrow}_{\phi_{m'}^{S}}+\ket{\uparrow}_{s.c}\otimes\ket{\downarrow}_{\phi_{m'}^{S}}}^\textrm{Entangled state}\\
&&\mbox{No flip:}\,\,\,\,\,\, \overbrace{\ket{\downarrow}_{s.c}}^\textrm{Josephson current state}\times\overbrace{\ket{\downarrow}_{\phi_{m'}^{S}}}^\textrm{Spin-flipper state}\xrightarrow{\textrm{interaction}}\overbrace{\ket{\downarrow}_{s.c}\times\ket{\downarrow}_{\phi_{m'}^{S}}}^\textrm{Product state}.
\end{eqnarray}
\end{widetext}
In our case spin-flipper interacts with Josephson spin current state. The whole macroscopic wavefunction of supercurrent is entangled with spin flipper wavefunction.}


\begin{thebibliography}{99}
\bibitem{hei}F. Giazotto, T. T. Heikkil\"{a}, G. P. Pepe, P. Helist\"{o}, A. Luukanen, J. P. Pekola, Ultrasensitive proximity
Josephson sensor with kinetic inductance readout, Appl. Phys. Lett. 92, 162507 (2008).
\bibitem{gua}C. Guarcello, A. Braggio, P. Solinas, G. P. Pepe, F. Giazotto, Josephson-Threshold Calorimeter, Phys. Rev. Applied 11, 054074 (2019). 
\bibitem{pso}P. Solinas, F. Giazotto, G. P. Pepe, Proximity SQUID Single-Photon Detector via Temperature-to-Voltage Conversion, Phys. Rev. Applied 10, 024015 (2018).
\bibitem{zgi}M. Zgirski, M. Foltyn, A. Savin, K. Norowski, M. Meschke, J. Pekola, Nanosecond Thermometry with Josephson Junctions, Phys. Rev. Applied 10, 044068 (2018).
\bibitem{wang}L. B. Wang, O. P. Saira, J. P. Pekola, Fast thermometry with a proximity Josephson junction, Appl. Phys. Lett. 112, 013105 (2018).
\bibitem{luu}F. Giazotto, T. T. Heikkil\"{a}, A. Luukanen, A. M. Savin, J. P. Pekola, Opportunities for mesoscopics in thermometry and refrigeration: Physics and applications, Rev. Mod. Phys. 78, 217 (2006).
\bibitem{yan}J. Yang, J, C. Elouard, J. Splettstoesser, B. Sothmann, R. S\'anchez, A. N. Jordan, Thermal transistor and
thermometer based on Coulomb-coupled conductors, Phys. Rev. B 100, 045418 (2019).
\bibitem{soth}B. Sothmann, F. Giazotto, E. M. Hankiewicz, High-efficiency thermal switch based on topological Josephson junctions, New J. Phys. 19, 023056 (2017).
\bibitem{dutta}B. Dutta et al., Thermal Conductance of a Single-Electron Transistor, Phys. Rev. Lett. 119, 077701 (2017).
\bibitem{mar}G. Marchegiani, P. Virtanen, F. Giazotto, and M. Campisi, Self-Oscillating Josephson Quantum Heat Engine, Phys. Rev. Applied 6, 054014 (2016).
\bibitem{vis}F. Vischi, M. Carrega, P. Virtanen, E. Strambini, A. Braggio, F. Giazotto, Thermodynamic cycles in Josephson junctions, Sci Rep 9, 3238 (2019).
\bibitem{kar}B. Karimi and J. P. Pekola, Otto refrigerator based on a superconducting qubit: Classical and quantum performance, Phys. Rev. B 94, 184503 (2016).
\bibitem{mml} M. M. Leivo, J. P. Pekola, D. V. Averin, Efficient Peltier refrigeration by a pair of normal metal$/$insulator$/$superconductor junctions, Appl. Phys. Lett. 68, 1996 (1996).
\bibitem{ngu} H. Q. Nguyen, J. T. Peltonen, M. Meschke, and J. P. Pekola, Cascade Electronic Refrigerator Using Superconducting Tunnel Junctions, Phys. Rev. Applied 6, 054011 (2016).
\bibitem{bos} P. Solinas, R. Bosisio, and F. Giazotto, Microwave quantum refrigeration based on the Josephson effect, Phys. Rev. B 93, 224521 (2016).
\bibitem{gma} G. Marchegiani, P. Virtanen and F. Giazotto, On-chip cooling by heating with superconducting tunnel junctions, Europhys. Lett. 124, 48005 (2018).
\bibitem{huss} R. Hussein et al., Nonlocal thermoelectricity in a Cooper-pair splitter, Phys. Rev. B 99, 075429 (2019).
\bibitem{bra} G. Marchegiani, A. Braggio, and F. Giazotto, Nonlinear Thermoelectricity with Electron-Hole Symmetric Systems, Phys. Rev. Lett. 124, 106801 (2020).
\bibitem{rko} R. Kosloff, Quantum Thermodynamics: A Dynamical Viewpoint, Entropy 15, 2100-2128 (2013).
\bibitem{kos} R. Kosloff and A. Levy, Quantum heat engines and refrigerators: Continuous devices, Annu. Rev. Phys. Chem. 65, 365 (2014).
\bibitem{mjm} F. Giazotto and M. J. Martinez-Perez, The Josephson heat interferometer, Nature 492, 401 (2012).
\bibitem{est} P. Virtanen, F. Vischi, E. Strambini, M. Carrega, and F. Giazotto, Quasiparticle entropy in superconductor/normal metal/superconductor proximity junctions in the diffusive limit, Phys. Rev. B 96, 245311 (2017).
\bibitem{for} A. Fornieri et al., Nanoscale phase engineering of thermal transport with a Josephson heat modulator, Nat. Nanotechnol. 11, 258 (2016).
\bibitem{afo} A. Fornieri and F. Giazotto, Towards phase-coherent caloritronics in superconducting circuits, Nat. Nanotechnol. 12, 944 (2017).
\bibitem{bsc} B. Scharf et al., Topological Josephson Heat Engine, Commun Phys 3, 198 (2020).
\bibitem{carr} F. Vischi, M. Carrega, A. Braggio, P. Virtanen, and F. Giazotto, Thermodynamics of a Phase-Driven Proximity
Josephson Junction, Entropy 21, 1005 (2019).
\bibitem{butt}M. B\"{u}ttiker and T. M. Klapwijk, Flux sensitivity of a piecewise normal and superconducting metal loop, Phys. Rev. B 33, 5114(R) (1986).
\bibitem{BTK}G. E. Blonder, M. Tinkham and T. M. Klapwijk, Transition from metallic to tunneling regimes in superconducting
microconstrictions: Excess current, charge imbalance, and supercurrent conversion, Phys. Rev. B 25, 4515 (1982).
\bibitem{AJP}O. L. T de Menezes and J. S Helman, Spin flip enhancement at resonant transmission, Am. J. Phys 53, 1100 (1985).
{\bibitem{ASC}A. Mani, S. Pal and C. Benjamin, Designing a highly efficient graphene quantum spin heat engine, Sci Rep 9, 6018 (2019).}
\bibitem{Maru}G. Cordourier-Maruri, Y. Omar, R. de Coss, and S. Bose, Graphene-enabled low-control quantum gates between static and mobile spins,  Phys. Rev. B 89, 075426 (2014).
\bibitem{Liu}H. D. Liu, X. X. Yi, Geometric phases in a scattering process, Phys. Rev. A 84, 022114 (2011).
\bibitem{FC}F. Ciccarello, G. M. Palma, and M. Zarcone, Entanglement-induced electron coherence in a mesoscopic ring with two magnetic impurities, Phys. Rev. B 75, 205415 (2007).
\bibitem{ysr}S. Pal and C. Benjamin, Yu-Shiba-Rusinov bound states induced by a spin flipper in the vicinity of a s-wave superconductor, Scientific Reports  8: 11949 (2018).
\bibitem{LINDER}H. Enoksen, J. Linder and A. Sudb\o{}, Spin-flip scattering and critical currents in ballistic half-metallic d-wave Josephson junctions, Phys. Rev. B 85,014512 (2012).
\bibitem{Kri} A. Krichevsky, M. Schechter, Y. Imry, \& Y. Levinson, Spectrum and thermodynamic currents in one-dimensional Josephson elements, Phys. Rev. B 61, 3723 (2000).
\bibitem{annu}G. Annunziata, H. Enoksen, J. Linder, M. Cuoco, C. Noce and A. Sudb\o{}, Josephson effect in S/F/S junctions: Spin bandwidth asymmetry versus Stoner exchange, Phys. Rev. B 83, 144520 (2011).
\bibitem{Been} C. W. J. Beenakker, Universal limit of critical-current fluctuations in mesoscopic Josephson junctions, Phys. Rev. Lett. 67, 3836 (1991).
{\bibitem{kulik}I. O. Kulik, Macroscopic quantization and the proximity effect in s-n-s junctions, Sov. Phys. JETP 30, 944 (1970).}
\bibitem{golu} A. A. Golubov, M. Y. Kupriyanov, and E. II'ichev, The current-phase relation in Josephson junctions, Rev. Mod. Phys. 76, 411 (2004).
\bibitem{deG}P. G. de Gennes, Superconductivity of Metals and Alloys (Benjamin, New York, 1966).
\bibitem{biz}J. P. S. Bizarro, Comment on \textquotedblleft Not all counterclockwise thermodynamic cycles are refrigerators\textquotedblright [Am. J. Phys. 84, 413-418 (2016)], Am. J. Phys. 85, 861-863 (2017).
{\bibitem{amado}M. Amado, A. Fornieri, G. Biasiol, L. Sorba, \& F. Giazotto, A ballistic two-dimensional-electron-gas andreev interferometer, Applied Physics Letters 104, 242604 (2014).}
\bibitem{iso} I. Sochnikov et al., Nonsinusoidal Current-Phase Relationship in Josephson Junctions from the 3D Topological Insulator HgTe, Phys. Rev. Lett. 114, 066801 (2015).
\bibitem{sha}S. Hart et al., Current-phase relations of InAs nanowire Josephson junctions: From interacting to multimode regimes, Phys. Rev. B 100, 064523 (2019).
\bibitem{iko} I. Sochnikov et al., Direct Measurement of Current-Phase Relations in Superconductor/Topological Insulator/Superconductor Junctions, Nano Lett. 13, 3086 (2013).
\end{thebibliography}
\end{document}